\documentclass[aps,prd,groupedaddress,showpacs]{revtex4}
\usepackage{epsfig}
\usepackage{latexsym}

\begin{document}

\title{
Hyperbolic triangle in the special theory of relativity}
\author{Yongkyu Ko}
\email[]{yongkyu@phya.yonsei.ac.kr}
\homepage[]{http://phya.yonsei.ac.kr/~yongkyu}
\affiliation{
Department of Physics, Yonsei University, Seoul 120-749, Korea}
\date{\today}

\begin{abstract}
The vector form of a Lorentz transformation which is separated with time
and space parts is studied.  It is necessary to introduce a new definition
of the relative velocity in this transformation, which plays an important
role for the calculations of various invariant physical
quantities.  The Lorentz transformation expressed with this vector form is
geometrically well interpreted in a hyperbolic space.
It is shown that the Lorentz transformation can be interpreted as the law
of cosines and sines for a hyperbolic triangle in hyperbolic trigonometry.
So the triangle made by the two origins of inertial frames and a moving
particle has the angles whose sum is less than $180~^o$.
\end{abstract}

\pacs{02.40Yy, 03.30+p, 11.30.Cp, 45.20.-d, 98.80Hw}
\maketitle

\section{Introduction}
The special theory of relativity starts with the fundamental two postulates
which are for relativity and the constancy of the speed of light in all
inertial frames \cite{Goldstein,Jackson,Marion}.
The transformation between inertial frames is known
as the Lorentz transformation which satisfies the two postulates.  Therefore
physical quantities and formulas are the same forms in all inertial frames.
This is achieved by using the four vector notation which is combined with the
space and time components, such as $v^{\mu}=dx^{\mu}/d \tau$,
$F^{\mu}=dp^{\mu}/d \tau$ and so on \cite{Misner}.
However the separate notation of
four vectors does not show the covariance manifestly in the transformation
equations.  The transformed
coordinate system used in the standard texts \cite{Goldstein,Jackson}
has a little complicated transformation formulas, such as,
\begin{eqnarray}
t' &=& \gamma (t - \mbox{\boldmath{$v$}} \cdot \mbox{\boldmath{$r$}}),
\nonumber \\
\mbox{\boldmath{$r$}}' &=& \mbox{\boldmath{$r$}} + {\gamma -1 \over{v^2}}
(\mbox{\boldmath{$v$}} \cdot \mbox{\boldmath{$r$}})
\mbox{\boldmath{$v$}} - \gamma \mbox{\boldmath{$v$}} t, \label{first}
\end{eqnarray}
thus, it is often doubtful that the transformed coordinate system has the same
physical formulas as the original coordinate system has.  These transformation
equations are not true vector forms.  They are merely components of the
vector equation, because the basis vectors are used in common in the two frames.
Physical quantities are measured in an inertial frame with its own units of
time and length.
Physical formulas composed of these quantities which are scalars, vectors or
tensors are formed in a frame with its own unit vectors.  Therefore it is
difficult to say relativity in the two frames with the above equations, because
the transformed equations often suffer from unwanted factors, such as, the
$\gamma$ factor.
Due to this trouble maker, the Biot-Savart law, for an example in
electromagnetism, is explained with sophisticated terms, such as, the duration
time of measurement for the magnetic field come from the transformed electric
field of a moving charged particle \cite{Jackson}.

A rotating coordinate system is a good example for the above arguments, of which
transformation is
\begin{equation}
x'_i=[ e^{i J_k \theta_k} ]_{ij}x_j,
\end{equation}
where $J_k$ is a generator of the rotation group $O(3)$.
Its time derivative is the velocity, which can be calculated with the
transformation equation as follows
\begin{eqnarray}
v'_i &=& {d x'_i \over{dt}} = [ e^{i J_k \theta_k} ]_{ij} ({d x_j \over{dt}}
+ i [J_k]_{jl}{d \theta_k \over{dt}}x_l) \nonumber \\
&=& [ e^{i J_k \theta_k} ]_{ij}( v_j + \epsilon_{jkl} \omega_k x_l)
\end{eqnarray}
where $[J_k]_{jl} = i \epsilon_{jkl}$ in the regular representation of the
rotation group \cite{Close}.  Then the vector form of the velocity is
\begin{eqnarray}
\mbox{\boldmath{$v$}}' &=& v'_i \hat{e}'_i = [ e^{i J_k \theta_k} ]_{ij}
( v_j + \epsilon_{jkl} \omega_k x_l) \hat{e}'_i \nonumber \\
&=& ( v_j + \epsilon_{jkl} \omega_k x_l) \hat{e}_j \nonumber \\
&=& \mbox{\boldmath{$v$}} + \mbox{\boldmath{$ \omega$}}
\times\mbox{\boldmath{$r$}},
\end{eqnarray}
where the unprimed basis vectors are transformed as $\hat{e}_j = \hat{e}'_i
[ e^{i J_k \theta_k} ]_{ij}$.
The acceleration is
\begin{equation}
a'_i = {d v'_i \over{dt}}
= [ e^{i J_k \theta_k} ]_{ij}( a_j + \epsilon_{jkl} \omega_k \epsilon_{lmn}
\omega_m x_n + 2 \epsilon_{jkl} \omega_k v_l), \label{accel}
\end{equation}
where we consider a constant angular velocity.
The vector form of the above equation is
\begin{equation}
\mbox{\boldmath{$a$}}' = \mbox{\boldmath{$a$}} + \mbox{\boldmath{$\omega$}}
\times
(\mbox{\boldmath{$\omega$}} \times \mbox{\boldmath{$r$}}) +
2 \mbox{\boldmath{$\omega$}} \times \mbox{\boldmath{$v$}},
\end{equation}
and, if a mass is multiplied to the equation, the Newton's law is
\begin{equation}
\mbox{\boldmath{$F$}}' = m\mbox{\boldmath{$a$}}'= m\mbox{\boldmath{$a$}}
+ m\mbox{\boldmath{$\omega$}} \times
(\mbox{\boldmath{$\omega$}} \times \mbox{\boldmath{$r$}})
+ 2 m\mbox{\boldmath{$\omega$}} \times \mbox{\boldmath{$v$}},\label{newton}
\end{equation}
where the second term is the centrifugal force and the last term is the
Coriolis force as shown in the standard
texts \cite{Goldstein,Marion}.
Therefore the position vectors in the two frames are written by
\begin{equation}
\mbox{\boldmath{$r$}}' = x'_i \hat{e}'_i
= [ e^{i J_k \theta_k} ]_{ij}x_j \hat{e}'_i
=  x_j \hat{e}_j =\mbox{\boldmath{$r$}},
\end{equation}
irrespective of the rotation angle which is constant or varying with time.
Since the mass times acceleration is expressed as the same forms in the two
frames, the Newton's law can be said to be expressed as covariant
manner in the
vector representation rather than in the component representation of the vector.
There is no rotation matrix, namely, an unwanted factor, in Eq. (\ref{newton})
compared to Eq. (\ref{accel}).

Therefore the correct vector form of the Lorentz transformation should be
expressed with its own components and unit vectors.  It is more natural that
relativity can be expected after obtaining the correct vector form of the
Lorentz transformation.
As learned from non-relativistic kinematics, if a physical phenomenon
is described with a geometrical picture, it is much easier to comprehend it.
The correct vector forms of the Lorentz transformation are well interpreted
with geometrical pictures, which are the law of cosines and sines in
hyperbolic trigonometry.  It is of course already known that the velocity space
of the special theory of relativity is a Lobatchewsky space, that is, a
hyperbolic space \cite{Laptev}, but our interpretation gives more clear
extended explanation for the hyperbolic space.
So the Thomas precession can be interpreted as the
time rate of the angle defect.

The outline of this paper is that a rotational transformation is adapted to
a Lorentz transformation in section II in order to obtain the correct vector
form of the Lorentz transformation, the method of
calculation for invariant quantities under the Lorentz transformation and the
geometrical interpretations of the Lorentz transformation.  In section III, most
of basic physical quantities are investigated under the Lorentz transformation
according to the guidelines of section II, and developed further.  In
section IV, the Lorentz transformation for an energy and a momentum is shown to
be the law of cosines and sines in hyperbolic trigonometry.  Finally some
conclusions are given.  Since spherical and hyperbolic trigonometries are not
seen in the standard texts for physics,
they are introduced in Appendices A and B.
Spherical trigonometry is very useful to introduce an angle vector as shown in
Appendix A and to comprehend hyperbolic trigonometry and its space.

\section{Rotational transformation}
\begin{figure}
\centerline{\epsfig{file=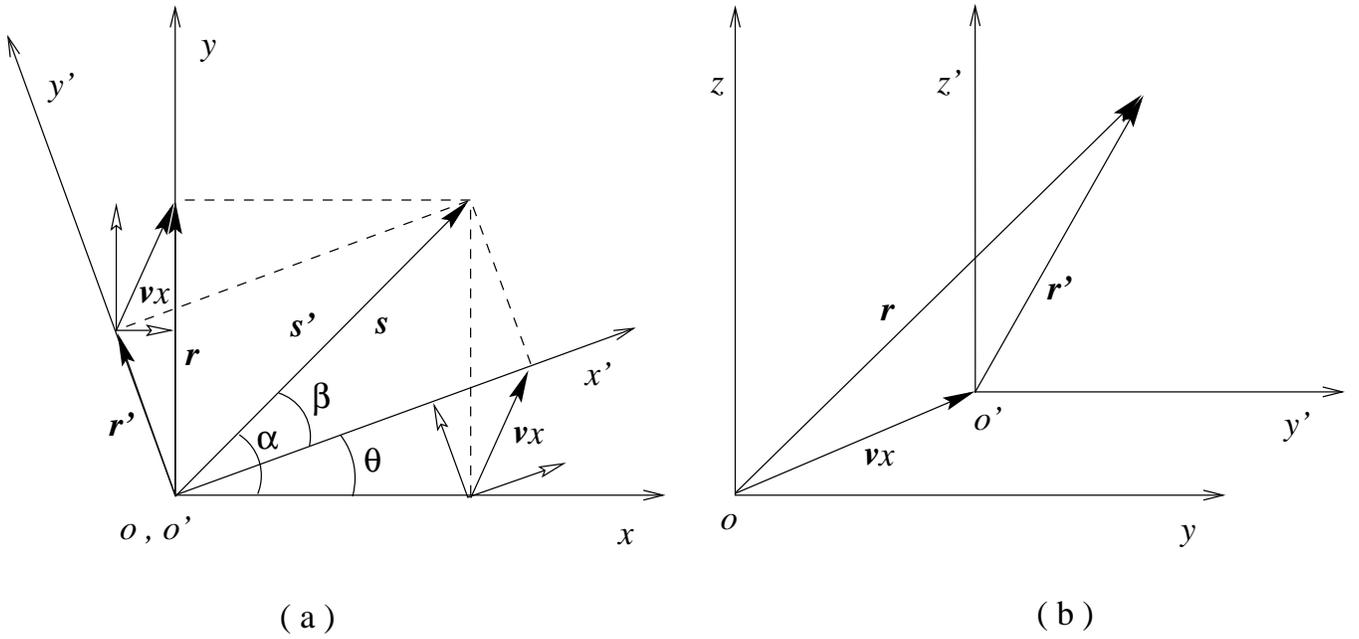, width=1.0\columnwidth}}
\caption{The rotational transformation (a) around $z$-axis can be analogized
to the Lorentz transformation as shown in the figure (b) by regarding
as $\tan \theta =v$.  The new relative velocity vector can be decomposed into
the two vectors depicted with unfilled arrows in the primed coordinate system
and unprimed coordinate system, respectively. }
\label{fig1}
\end{figure}

A Lorentz transformation is similar to a rotational transformation,
if the relative velocity $v$ between two frames is regarded as tangent of
a rotational angle $\phi$, that is, $\tan \phi=-iv$.  So the converse of the
similarity is also true and gives us a good insight for the special theory
of relativity.  Moreover they are expressed as the same transformation formulas
in the Euclidean space time.
A rotated coordinate system $K'$ with an angle $\theta$
around the z-axis of a coordinate system $K$ as shown in Fig. \ref{fig1}
has the components as follows
\begin{eqnarray}
x'&=& x \cos \theta + y \sin \theta, \nonumber \\
y'&=& - x \sin \theta + y \cos \theta, \nonumber \\
z' &=& z.
\end{eqnarray}
Then let $\cos \theta$ and $\sin \theta$ be assigned to $1/\sqrt{1+v^2}$ and
$v/\sqrt{1+v^2}$, respectively.  The above transformation becomes similar to
a Lorentz transformation as follows
\begin{eqnarray}
x' &=& {1 \over{\sqrt{1+v^2}}}(x + v y) = \gamma (x + v y), \nonumber \\
y' &=& {1 \over{\sqrt{1+v^2}}}(y - v x) = \gamma (y - v x), \nonumber \\
z' &=& z,
\end{eqnarray}
where $\gamma$ is used as $1/\sqrt{1+v^2}$ only in this section.
We shall see in the next section that the difference between the rotational
and the Lorentz transformations comes from that between the metrics in the
two transformations, if the coordinate $x$ is regarded as
time axis and the coordinates $y$ and $z$ are done as space axes.
Hence the signs before the terms involving $v$ are
different from those of the Lorentz transformation.

The vector forms of the Lorentz transformation used in the standard
texts\cite{Goldstein,Jackson} are again written in the version of this
rotational transformation by
\begin{eqnarray}
x' &=& \gamma (x + v y) = \gamma (x + \mbox{\boldmath{$v$}} \cdot
\mbox{\boldmath{$r$}}), \nonumber \\
\mbox{\boldmath{$r$}}' &=& y' \hat{j}' + z' \hat{k} ' \nonumber \\
 &=& y \hat{j}' + z \hat{k}' + (\gamma-1) y \hat{j}' - \gamma v x \hat{j}'
 \nonumber \\
 &=& \mbox{\boldmath{$r$}} + {\gamma -1 \over{v^2}} (\mbox{\boldmath{$r$}}
 \cdot \mbox{\boldmath{$v$}}) \mbox{\boldmath{$v$}} - \gamma
 \mbox{\boldmath{$v$}} x, \label{vector1}
\end{eqnarray}
where the basis vectors are used in common in the two coordinate systems.
These equations are not, therefore, true vector forms, but tell us only
components of the transformation equation.
The correct definition of the vector in the unprimed coordinate system
should be
\begin{equation}
\mbox{\boldmath{$r$}} = y \hat{j} + z \hat{k}.
\end{equation}
Hence the following transformation for the basis vectors should be applied to
the above equation (\ref{vector1}):
\begin{eqnarray}
\hat{i}'&=& \hat{i} \cos \theta + \hat{j} \sin \theta, \nonumber \\
\hat{j}'&=& - \hat{i} \sin \theta + \hat{j} \cos \theta, \nonumber \\
\hat{k}' &=& \hat{k}.
\end{eqnarray}
Then the vector form of the transformation equation is calculated as
\begin{eqnarray}
\mbox{\boldmath{$r$}}'
&=& \mbox{\boldmath{$r$}} + y' \hat{j}' - y \hat{j} \nonumber \\
 &=& \mbox{\boldmath{$r$}} - {y \hat{j} - y' \hat{j}' \over{x}} x \nonumber \\
 &=& \mbox{\boldmath{$r$}} - {\bf v} x, \label{vel1}
\end{eqnarray}
where ${\bf v}$ is defined as a new relative velocity between the two frames
at the point indicated by the coordinates in the space time,
and used a different type letter to distinguish the new relative velocity
vector
from usual vectors.  We will call this type velocity the new relative velocity
from now on because we have no name to call it suitably.
The vectors in the two frames are connected via this new relative velocity.
The definition of the new relative velocity is exactly correct in a sense that,
if we consider it in general situation, namely, curvilinear coordinate
systems, the variation of a vector should take into account both its component
and unit vector, such as, $y \hat{j}-y' \hat{j}'=y \hat{j} - y' \hat{j}
+ y' \hat{j} - y' \hat{j}' = \Delta y \hat{j} + y' \Delta \hat{j}$.
Therefore the definition of a velocity should be
\begin{equation}
{\bf v} = {d{\bf r} \over dt} = {dr {\hat{\bf
r}}+ r d{\hat{\bf r}} \over{dt}},\label{genvel}
\end{equation}
even in the rectangular coordinate system, because the basis vectors are
defined only in an inertial frame.
The $x$ coordinate has the following transformation:
\begin{eqnarray}
\mbox{\boldmath{$x$}}' &=& \mbox{\boldmath{$x$}} + x' \hat{i}' - x \hat{i}
= \mbox{\boldmath{$x$}} + {y \hat{j} - y' \hat{j}' \over{x}}x \nonumber \\
&=& \mbox{\boldmath{$x$}} + {\bf v} x, \label{tranx}
\end{eqnarray}
because the vectors $\mbox{\boldmath{$s$}}$ and $\mbox{\boldmath{$s$}}'$
in the two coordinate systems are identical as follows
\begin{equation}
\mbox{\boldmath{$s$}}' = x' \hat{i}' + y' \hat{j}' + z' \hat{k}'
= x \hat{i} + y \hat{j} + z \hat{k} = \mbox{\boldmath{$s$}},
\end{equation}
and we get the following equation:
\begin{equation}
x' \hat{i}' - x \hat{i} = y \hat{j}- y' \hat{j}'.
\end{equation}
The component forms of the Lorentz transformation to which we are accustomed
can be calculated by multiplying the basis vectors to Eqs. (\ref{vel1}) and
(\ref{tranx}) as
\begin{eqnarray}
x' &=& \mbox{\boldmath{$x$}}' \cdot \hat{i}' = x \hat{i} \cdot \hat{i}' +
{\bf v} \cdot \hat{i}' x=\gamma x + {\bf v}'
\cdot \mbox{\boldmath{$r$}} = x \cos \theta + y \sin \theta,  \nonumber \\
y' &=& \mbox{\boldmath{$r$}}' \cdot \hat{j}' = y \hat{j} \cdot \hat{j}' -
{\bf v} \cdot
\hat{j}' x = y \cos \theta - x \sin \theta, \nonumber \\
z' &=& \mbox{\boldmath{$r$}}' \cdot \hat{k}' = z \hat{k} \cdot \hat{k}' -
{\bf v} \cdot \hat{k}' x = z,
\end{eqnarray}
where the scalar products of the basis vectors are the direction cosines of each
axis and the scalar products of the new velocity vector to the basis vectors
are calculated as
\begin{eqnarray}
{\bf v} \cdot \hat{i}' &=& {(y \hat{j} - y' \hat{j}') \cdot \hat{i}' \over{x}}
= {y \hat{j}\cdot \hat{i}' \over{x}} = {y \sin \theta \over{x}}, \nonumber \\
{\bf v} \cdot \hat{j}' &=& {(y \hat{j} - y' \hat{j}') \cdot \hat{j}' \over{x}}
= {y \hat{j}\cdot \hat{j}' -y' \over{x}} = \sin \theta, \nonumber \\
{\bf v} \cdot \hat{k}' &=& {(y \hat{j} - y' \hat{j}') \cdot
\hat{k}' \over{x}} = 0.
\end{eqnarray}

These equations tell us the direction of the new velocity vector as shown in
Fig.
\ref{fig1}, the new velocity vector is represented in the primed and unprimed
coordinate systems as
\begin{eqnarray}
{\bf v} x &=& y' \sin \theta \hat{i} + x' \sin \theta \hat{j} \nonumber \\
&=& y \sin \theta \hat{i}' + x \sin \theta \hat{j}' = {\bf v}' x'.
\end{eqnarray}
If the new relative velocity in the primed coordinate system is defined as
this equation, the vector forms of the transformation equations
(\ref{vel1})
and (\ref{tranx}) are expressed as covariant forms in the two frames.
The scalar product of the new relative velocity to the position vector is
calculated as
\begin{eqnarray}
{\bf v} \cdot \mbox{\boldmath{$r$}} &=& {1 \over{x}}(y \hat{j} - y' \hat{j}')
\cdot(y \hat{j} + z \hat{k}) \nonumber \\
&=& {1 \over{x}}(y^2-y' y\cos \theta) = {1 \over{x}}(y^2 \sin^2 \theta +x y
\sin \theta \cos \theta ) \nonumber \\
&=& {x' \over{x}} y \sin \theta = x' {\bf v} \cdot \hat{i}'.
\label{vdotr1}
\end{eqnarray}
The new relative velocity in the primed coordinate
system can be defined as
\begin{equation}
{\bf v}' = {y \hat{j} - y' \hat{j}' \over{x'}},\label{vel2}
\end{equation}
and the scalar product of it to the position vector is related with the above
relevant quantities as follows
\begin{equation}
{\bf v}' \cdot \mbox{\boldmath{$r$}} = x {\bf v} \cdot
\hat{i}' = y \sin \theta ={x \over{x'}} {\bf v} \cdot \mbox{\boldmath{$r$}}.
\label{vdotr2}
\end{equation}
The new relative velocities ${\bf v}$ and ${\bf v}'$ are defined different from
each other in general in the two coordinate systems.  However the new relative
velocities of the origins of each other frames are reduced to the usual vectors
\begin{equation}
{\bf v} = {y \over x} \hat{j} = v \hat{j} = \mbox{\boldmath{$v$}},
~~~{\bf v}' = -{y' \over x'} \hat{j}' = - v \hat{j}' = \mbox{\boldmath{$v$}}',
\end{equation}
from Eqs. (\ref{vel1}) and (\ref{vel2}), where the magnitude of the usual
relative velocities are the same in both frames.  This is assumed without
mention as a postulate in the special theory of relativity.  The usual
relative velocity vectors are different from each other in their direction
because their unit vectors are different.
In general all moving particles have their own
basis vectors of the inertial frame so that they should have the form of the
velocity like
Eq. (\ref{genvel}) or the new relative velocities.
However their velocity vectors can be expressed as usual vectors
in an observing frame, because the particles are assumed to be at the origins of
their own frames as shown in above equations.  The vector form of the Lorentz
transformation requires the different new relative velocity from point to point
so that the magnitude of it is also varying according to points.
The magnitude of the new relative velocity is calculated as
\begin{eqnarray}
{\bf v} \cdot {\bf v}
&=& {1 \over{x^2}}(y^2+y'^2-2y y' \cos \theta) \nonumber \\
&=& {1 \over{x^2}}(x^2+y^2)\sin^2 \theta
= {\sin^2 \theta \over{\cos^2 \alpha}} = {\rm v}^2,
\label{vdotv}
\end{eqnarray}
where $\alpha$ is the angle of the vector $\mbox{\boldmath{$s$}}$ making
with $x$-axis as
shown in Fig. \ref{fig1}.
The magnitude of the new relative velocity observed in
the primed coordinate system is written by
\begin{equation}
{\bf v}' \cdot {\bf v}' = {1 \over{x'^2}}(x^2+y^2)\sin^2 \theta
= {\sin^2 \theta \over{\cos^2 \beta}} = {\rm v}'^2,
\end{equation}
where $\beta$ is the angle of the vector $\mbox{\boldmath{$s$}}$ making
with $x'$-axis as
shown in Fig. \ref{fig1}.  Therefore the relation between the two new relative
velocities has the identity:
\begin{equation}
{\rm v}^2 \cos^2 \alpha = {\rm v}'^2 \cos \beta = \sin^2 \theta
= {v^2 \over{1+v^2}}.
\end{equation}
Now the vector $\mbox{\boldmath{$s$}}$ is invariant under the rotational
transformation so
that the transformed vector $\mbox{\boldmath{$s$}}'$ is calculated as
\begin{eqnarray}
\mbox{\boldmath{$s$}}'^2 &=& x'^2 + \mbox{\boldmath{$r$}}' \cdot
\mbox{\boldmath{$r$}}' \nonumber \\
&=& \gamma^2 x^2 + 2 \gamma x {\bf v}' \cdot \mbox{\boldmath{$r$}}
+ ({\bf v}' \cdot \mbox{\boldmath{$r$}})^2
+\mbox{\boldmath{$r$}} \cdot \mbox{\boldmath{$r$}} - 2 x {\bf v} \cdot
\mbox{\boldmath{$r$}} + x^2 {\bf v \cdot v} \nonumber \\
&=& x^2 + \mbox{\boldmath{$r$}} \cdot \mbox{\boldmath{$r$}}
= \mbox{\boldmath{$s$}}^2, \label{invr}
\end{eqnarray}
where Eqs. (\ref{vdotr1}), (\ref{vdotr2}), and (\ref{vdotv}) are inserted in
the second line of the equation.  This calculation is a little different from
the calculation in the component form of the Lorentz transformation,
but leads to the
same result.

\section{Lorentz transformation}
The vector form of a Lorentz transformation for a coordinate system moving
with a velocity along the $x$-axis of a rest coordinate system is trivially
achieved with the transformation of the previous section by considering the
difference of the metrics.  This can be also
obtained from the general transformation below, let us pay attention to the
general case.
A Lorentz transformation for an arbitrary direction of the relative
velocity with respect to the origin of the rest frame is just Eq. (\ref{first})
whose matrix form is
\begin{equation}
\left(\begin{array}{c}
t' \\ x' \\ y' \\ z' \end{array} \right) = \left(\begin{array}{cccc}
\gamma & - \gamma v n_x & - \gamma v n_y &  - \gamma v n_z \\
- \gamma v n_x &
1+(\gamma - 1) n_x^2 & (\gamma - 1) n_x n_y & (\gamma - 1) n_x n_z \\
- \gamma v n_y &
(\gamma - 1) n_x n_y & 1 + (\gamma - 1) n_y^2 & (\gamma - 1) n_y n_z \\
- \gamma v n_z & (\gamma - 1) n_x n_z &
(\gamma - 1) n_y n_z & 1 + (\gamma - 1) n_z^2 \\ \end{array}
\right) \left(\begin{array}{c}
t \\ x \\ y \\ z \end{array} \right), \label{mat}
\end{equation}
which are the array of the components of Eq. (\ref{first})
\cite{Goldstein,Jackson}, and where $v$ is the usual relative velocity between
the origins of the two inertial frames.  The
$\gamma$, $\gamma v$ and $v$ are assigned to $\cosh \vartheta$, $\sinh
\vartheta$ and $\tanh \vartheta$, respectively, and the direction cosines
of the relative velocity vector can be thought of as
$n_x=\sin \theta \cos \phi$, $n_y=\sin \theta \sin \phi$, and $n_z=\cos \theta$.
The basis vectors of the coordinate system are also transformed as
\begin{equation}
\left(\begin{array}{c}
\hat{h}' \\ \hat{i}' \\ \hat{j}' \\ \hat{k}' \end{array} \right) =
\left(\begin{array}{cccc}
\gamma &  \gamma v n_x &  \gamma v n_y &   \gamma v n_z \\
\gamma v n_x &
1+(\gamma - 1) n_x^2 & (\gamma - 1) n_x n_y & (\gamma - 1) n_x n_z \\
\gamma v n_y &
(\gamma - 1) n_x n_y & 1 + (\gamma - 1) n_y^2 & (\gamma - 1) n_y n_z \\
\gamma v n_z & (\gamma - 1) n_x n_z &
(\gamma - 1) n_y n_z & 1 + (\gamma - 1) n_z^2 \\
\end{array} \right) \left(\begin{array}{c}
\hat{h} \\ \hat{i} \\ \hat{j} \\ \hat{k} \end{array} \right),
\end{equation}
where the transformation matrix is not only inverse to the matrix of the
coordinate transformation, but also a transformation matrix for the
corresponding covariant vector to the coordinate.

From the calculations according to the previous section, a position and a time
vectors in the moving frame are written by
\begin{eqnarray}
\mbox{\boldmath{$r$}}' &=& \mbox{\boldmath{$r$}} - {\bf v} t, \nonumber \\
\mbox{\boldmath{$t$}}' &=& \mbox{\boldmath{$t$}} + {\bf v} t, \label{coovec}
\end{eqnarray}
where the transformation for the position vector is ironically similar to
the Galilean transformation, but an exact Lorentz
transformation due to the new relative velocity.
The new relative velocities are defined as
\begin{eqnarray}
{\bf v} &=& {x \hat{i} +y \hat{j} + z \hat{k}-x' \hat{i}'
- y' \hat{j}' - z' \hat{k}' \over{t}}
= {t' \hat{h}' - t \hat{h} \over{t}}, \nonumber \\
{\bf v}' &=& {x \hat{i} +y \hat{j} + z \hat{k}-x' \hat{i}'
- y' \hat{j}' - z' \hat{k}' \over{t}'}
= {t' \hat{h}' - t \hat{h} \over{t}'},\label{conewrel}
\end{eqnarray}
since an event four vector $\mbox{\boldmath{$s$}}$ in the rest frame is
identical to $\mbox{\boldmath{$s$}}'$ in the moving frame as follows
\begin{eqnarray}
\mbox{\boldmath{$s$}} &=& x^{\mu} \hat{e}_{\mu} = t \hat{h} + x \hat{i}
+ y \hat{j} + z \hat{k} \nonumber \\
&=& t' \hat{h}' + x' \hat{i}' + y' \hat{j}' + z' \hat{k}'
= x'^{\mu} \hat{e'}_{\mu} = \mbox{\boldmath{$s$}}'.
\end{eqnarray}
It should be noted that $\hat{h} \cdot \hat{h} = 1$ and $\hat{i} \cdot \hat{i}
= \hat{j} \cdot \hat{j} = \hat{k} \cdot \hat{k} = - 1$, because the metric is
represented in terms of its basis vectors as $g_{\mu \nu} =
\hat{e}_{\mu} \cdot \hat{e}_{\nu}$ \cite{Sokolnikoff}.
The time component of the transformation equations is calculated from the time
vector equation as
\begin{equation}
t' = \mbox{\boldmath{$t$}}' \cdot \hat{h}'= (\mbox{\boldmath{$t$}}+ {\bf v} t)
\cdot \hat{h}' =\gamma t + {\bf v}' \cdot \mbox{\boldmath{$r$}}
\end{equation}
where the scalar product of the time component basis vector to the new
relative velocity vector
is calculated as
\begin{equation}
t {\bf v} \cdot \hat{h}' = \hat{n} \cdot
\mbox{\boldmath{$r$}} \sinh \vartheta = {t \over{t'}} {\bf v} \cdot
\mbox{\boldmath{$r$}} = {\bf v}' \cdot \mbox{\boldmath{$r$}}.
\end{equation}
The scalar products of the new relative velocity to the other basis vectors are
\begin{equation}
{\bf v} \cdot \hat{i}' = - n_x \sinh \vartheta,~~
{\bf v} \cdot \hat{j}' = - n_y \sinh \vartheta,~~
{\bf v} \cdot \hat{k}' = - n_z \sinh \vartheta.
\end{equation}
From the above knowledge, the component equations of the Lorentz transformation
can be obtained form the scalar product of every basis vector
to Eq. (\ref{coovec}),
which agree to Eq. (\ref{mat}) as shown in the previous section.

The square of the new relative velocity vector is calculated from the above
definition as
\begin{eqnarray}
{\bf v} \cdot {\bf v} &=& ({t' \hat{h}' - t \hat{h} \over{t}})^2 \nonumber \\
&=&(-1 + {(\hat{n} \cdot \mbox{\boldmath{$r$}})^2 \over{t^2}})\sinh^2
\vartheta,
\end{eqnarray}
which is the same form as shown in the previous section.
Now we can check the following invariant quantity with using the above
calculations according to the procedure of the previous section:
\begin{equation}
\mbox{\boldmath{$s$}}'^2 = t'^2 + \mbox{\boldmath{$r$}}' \cdot
\mbox{\boldmath{$r$}}'= t'^2 - r'^2 = t^2 - r^2 = t^2 + \mbox{\boldmath{$r$}}
\cdot \mbox{\boldmath{$r$}}=\mbox{\boldmath{$s$}}^2.
\end{equation}
This equation may be useful to explain the time dilation phenomenon with the
muon decay in elementary course of relativity.
The origins of the two inertial frames which are the frame of a muon rest
frame and the frame of an observer on the earth are coincided at the position
and
time that a muon is produced in the atmosphere by a cosmic ray.  We observe
the muon with a velocity
$v$, and it travels a distance $r=vt$.  Since the muon does not travel in the
muon rest frame, that is, $r'=0$, the time interval which the muon travels is
observed as $t = t'/\sqrt{1-v^2}$ on the earth.

An inverse Lorentz transformation can be obtained by applying the inverse matrix
of the transformation to the transformation matrix equation Eq. (\ref{mat}),
so the direction cosines of the
relative velocity between the origins of the two frames are used in common in
both frames.  The unit vector of the relative velocity in the primed coordinate
system is, therefore, calculated as
\begin{equation}
\hat{n}'= n_x \hat{i}' + n_y \hat{j}' + n_z \hat{k}'
=\hat{n} \cosh \vartheta + \hat{h} \sinh \vartheta,
\end{equation}
which joins the first line of the transformation matrix equation for the basis
vectors and forms a pair of the Lorentz transformation for the unit vectors
as follows
\begin{eqnarray}
\hat{h}' &=& \hat{h} \cosh \vartheta + \hat{n} \sinh \vartheta, \nonumber \\
\hat{n}' &=& \hat{n} \cosh \vartheta + \hat{h} \sinh \vartheta, \nonumber \\
\hat{m}' &=& \hat{m},
\end{eqnarray}
where two unit vectors of $\hat{m}$ which are orthogonal to $\hat{n}$ can be
chosen as the remaining space basis vectors.
Using the transformation matrix equations
for the coordinate and the unit vectors,
the scalar and vector products of the unit vector
of the relative velocity to the position vector in the primed
coordinate system are directly
calculated as
\begin{eqnarray}
& &\hat{n}' \cdot \mbox{\boldmath{$r$}}' = - n_x x' - n_y y' - n_z z' = t \sinh
\vartheta + \hat{n} \cdot \mbox{\boldmath{$r$}} \cosh \vartheta, \nonumber \\
& &\hat{n}' \times \mbox{\boldmath{$r$}}' = (n_y z' - n_z y') \hat{i}' + (n_z x'
- n_x z') \hat{j}' + (n_x y' - n_y x') \hat{k}' = \hat{n} \times
\mbox{\boldmath{$r$}}, \nonumber \\
& &\hat{n}' \times (\hat{n}' \times \mbox{\boldmath{$r$}}')
= - \hat{n}'(\hat{n}' \cdot \mbox{\boldmath{$r$}}') - \mbox{\boldmath{$r$}}'
= \hat{n} \times (\hat{n} \times \mbox{\boldmath{$r$}}),
\end{eqnarray}
where the vector product is not well defined in relativity.
Such an unsatisfactory definition may be due to a non-commutative property
between the operation of a vector product and the Lorentz transformation,
or insufficient definitions for the extended indices, $0, 1, 2, 3$ in the four
dimension for a vector
product.  Anyway the operation after the transformation needs a
transformation rule for the vector product itself, that is,
the symbol $\times$,
but the transformation after
the operation does not need to do so.  Therefore the latter is regarded as the
prescription for a vector product in this works.
Moreover scalar products are well defined in the four dimension.  Vector
products which can be expressed with scalar products, such as triple vector
products, does not matter in the Lorentz transformation after the operation of
vector products.

Now the corresponding coordinate transformation to the above transformation for
the unit vectors is
\begin{eqnarray}
& &t' = t \cosh \vartheta + \hat{n} \cdot \mbox{\boldmath{$r$}}
\sinh \vartheta, \nonumber \\
& &\hat{n}' \cdot \mbox{\boldmath{$r$}}' = t \sinh \vartheta
+ \hat{n} \cdot \mbox{\boldmath{$r$}} \cosh \vartheta, \nonumber \\
& &\hat{n}' \times (\hat{n}' \times \mbox{\boldmath{$r$}}')
= \hat{n} \times (\hat{n} \times \mbox{\boldmath{$r$}}), \label{usualvectr}
\end{eqnarray}
where the minus sign in the matrix of the Lorentz transformation is hidden
here in the scalar product due to the metric.  Since a vector is decomposed
into a transverse and a normal components with respect to a specified unit
vector, such as $\mbox{\boldmath{$r$}} = -(\hat{n} \cdot \mbox{\boldmath{$r$}})
\hat{n} - \hat{n} \times (\hat{n} \times \mbox{\boldmath{$r$}})$, the
transformation for $\hat{n}' \times \mbox{\boldmath{$r$}}'$ is discarded
because it is irrelevant to the vector $\mbox{\boldmath{$r$}}$.
This transformation for an arbitrary direction of a relative velocity
agrees to that for a relative velocity along $x$-axis which can be obtained by
substituting $\hat{n}=\hat{i}$.
The position vectors in the transformation equations are
expressed exactly because they have their own components and unit vectors in
their initial frames.
The new relative velocity is written by
\begin{eqnarray}
{\bf v} t &=& {\bf v}' t' = \hat{n} \cdot \mbox{\boldmath{$r$}}
\sinh \vartheta \hat{h}' + t \sinh \vartheta \hat{n}' \nonumber \\
&=& \hat{n}' \cdot \mbox{\boldmath{$r$}}'
\sinh \vartheta \hat{h} + t' \sinh \vartheta \hat{n},\label{ga}
\end{eqnarray}
which gives the following identity:
\begin{equation}
(\hat{n} \cdot \mbox{\boldmath{$r$}}) \hat{h}' + t \hat{n}'
= (\hat{n}' \cdot \mbox{\boldmath{$r$}}') \hat{h} + t' \hat{n},
\end{equation}
by eliminating $\sinh \vartheta$ in the above equation.
The scalar products of the unit vectors $\hat{h}$ and $\hat{n}$ to the identity
give again the above Lorentz transformation with the correct vectors, that is,
Eq. (\ref{usualvectr}).
Therefore the new relative velocity has the Lorentz transformation properties
between the two vectors $\mbox{\boldmath{$r$}}$ and $\mbox{\boldmath{$r$}}'$ in
the two frames in Eq. (\ref{coovec}).  The above transformation
Eq. (\ref{usualvectr}) can be also obtained form the scalar and vector products
of the primed unit vector of the relative velocity to the transformed position
vector
as follows
\begin{eqnarray}
& & \hat{n}' \cdot \mbox{\boldmath{$r$}}' = \hat{n}' \cdot (\mbox{\boldmath{$r$}}
- {\bf v} t) = t \sinh \vartheta
+ \hat{n} \cdot \mbox{\boldmath{$r$}} \cosh \vartheta, \nonumber \\
& &\hat{n}' \times (\hat{n}' \times \mbox{\boldmath{$r$}}')
= - \hat{n}'(\hat{n}' \cdot \mbox{\boldmath{$r$}}') - (\mbox{\boldmath{$r$}}
- {\bf v}t)
= \hat{n} \times (\hat{n} \times \mbox{\boldmath{$r$}}),
\end{eqnarray}
by using the above new relative velocity and the transformation for the unit
vector.

Since the position vector is a sort of a displacement vector,
an infinitesimal displacement
vector can be investigated in the same way as follows
\begin{eqnarray}
d\mbox{\boldmath{$r$}}' &=& d\mbox{\boldmath{$r$}} - {\bf v} dt, \nonumber \\
d\mbox{\boldmath{$t$}}' &=& d\mbox{\boldmath{$t$}} + {\bf v} dt, \label{veldis}
\end{eqnarray}
where the new relative velocity is defined in the same way, but different from Eq.
(\ref{ga}) as follows
\begin{eqnarray}
{\bf v} dt &=& {\bf v}' dt' = \hat{n} \cdot d\mbox{\boldmath{$r$}}
\sinh \vartheta \hat{h}' + dt \sinh \vartheta \hat{n}' \nonumber \\
&=& \hat{n}' \cdot d\mbox{\boldmath{$r$}}'
\sinh \vartheta \hat{h} + dt' \sinh \vartheta \hat{n}.
\end{eqnarray}
This transformation for the infinitesimal displacement vector is not the
derivative of the transformation equations for the position and time vectors,
because the
new relative velocities between the two transformations are different from each
other.  A new relative velocity is always defined in the way of
Eq. (\ref{conewrel}).
A new relative velocity is defined different not only from point to point,
but also from vector to vector.
The invariant quantity for the infinitesimal displacement vector is also
calculated in the same way of the above event vectors as follows
\begin{equation}
d\mbox{\boldmath{$s$}}'^2 = dt'^2 + d\mbox{\boldmath{$r$}}' \cdot
d\mbox{\boldmath{$r$}}'= dt'^2 - dr'^2 = dt^2 - dr^2 = dt^2 + d\mbox{\boldmath{$r$}}
\cdot d\mbox{\boldmath{$r$}}=d\mbox{\boldmath{$s$}}^2.
\end{equation}
The transformation with usual vectors for the infinitesimal displacement vector
is represented as
\begin{eqnarray}
& & dt' = dt \cosh \vartheta + \hat{n} \cdot d\mbox{\boldmath{$r$}}
\sinh \vartheta, \nonumber \\
& & \hat{n}' \cdot d\mbox{\boldmath{$r$}}' = dt \sinh \vartheta
+ \hat{n} \cdot d\mbox{\boldmath{$r$}} \cosh \vartheta, \nonumber \\
& & \hat{n}' \times (\hat{n}' \times d\mbox{\boldmath{$r$}}')
= \hat{n} \times (\hat{n} \times d\mbox{\boldmath{$r$}}),
\end{eqnarray}
which can be used to calculate the transformation for the velocities of a moving
particle observed in both frames.
The components of the velocity of the particle which are parallel and
perpendicular  to the relative velocity are written by
\begin{eqnarray}
& & \hat{n}' \cdot \mbox{\boldmath{$u$}}' = {\hat{n} \cdot
(\mbox{\boldmath{$u$}} - \mbox{\boldmath{$v$}}) \over{1+\mbox{\boldmath{$v$}}
\cdot \mbox{\boldmath{$u$}}}}, \nonumber \\
& & \hat{n}' \times (\hat{n}' \times \mbox{\boldmath{$u$}}') = {\hat{n} \times
(\hat{n} \times \mbox{\boldmath{$u$}}) \over{\gamma (1+\mbox{\boldmath{$v$}}
\cdot \mbox{\boldmath{$u$}}})}, \label{veladdition}
\end{eqnarray}
by dividing the last two equations by $dt'$.
Using the above vector
equations for the infinitesimal displacement vector,
the velocity addition rule in vector form is written by
\begin{eqnarray}
\mbox{\boldmath{$u$}}' &=& {d{\mbox{\boldmath{$r$}}'} \over{dt'}}
={d\mbox{\boldmath{$r$}} - {\bf v}dt
\over{\gamma dt + {\bf v}' \cdot d {\bf r}}}\nonumber \\
&=& {\mbox{\boldmath{$u$}} - {\bf v} \over{\gamma + {\bf v}'
\cdot \mbox{\boldmath{$u$}}}} = {\mbox{\boldmath{$u$}} - {\bf v}
\over{\gamma (1+\mbox{\boldmath{$v$}} \cdot \mbox{\boldmath{$u$}})}},
\label{vel}
\end{eqnarray}
which is consistent with the above component equations by doing the scalar
and vector
products of the primed unit vector of the relative velocity to it.

Since an energy and a momentum are transformed as coordinates  under the
Lorentz transformation, the transformed momentum and energy vectors are
written by
\begin{eqnarray}
\mbox{\boldmath{$p$}}' &=& \mbox{\boldmath{$p$}} - {\bf v} E, \nonumber \\
\mbox{\boldmath{$E$}}' &=& \mbox{\boldmath{$E$}} + {\bf v} E, \label{mo}
\end{eqnarray}
where the new relative velocities in momentum space are also defined as
\begin{equation}
{\bf v} = {E' \hat{h}' - E \hat{h} \over{E}},~~{\bf v}'
= {E' \hat{h}' - E \hat{h} \over{E'}}.
\end{equation}
These velocities are calculated as
\begin{eqnarray}
{\bf v} E &=& {\bf v}' E' = \hat{n} \cdot \mbox{\boldmath{$p$}}
\sinh \vartheta \hat{h}' + E \sinh \vartheta \hat{n}' \nonumber \\
&=& \hat{n}' \cdot \mbox{\boldmath{$p$}}'
\sinh \vartheta \hat{h} + E' \sinh \vartheta \hat{n}.
\end{eqnarray}
These new relative velocities ${\bf v}$ and ${\bf v'}$ defined in momentum
space are equal
to those defined in the transformation for the infinitesimal displacement
vectors, if the velocity of the particle involved in the transformation is
expressed as $\mbox{\boldmath{$u$}} = d \mbox{\boldmath{$r$}}/dt
= \mbox{\boldmath{$p$}}/E$.
So the velocity addition rule is again written in vector form by
\begin{eqnarray}
\mbox{\boldmath{$u$}}' &=& {\mbox{\boldmath{$p$}}' \over{E'}}
={\mbox{\boldmath{$p$}} - {\bf v}E \over{\gamma E -
{\bf v}' \cdot \mbox{\boldmath{$p$}}}}\nonumber \\
&=& {\mbox{\boldmath{$u$}} - {\bf v} \over{\gamma + {\bf v}'
\cdot \mbox{\boldmath{$u$}}}},
\end{eqnarray}
which agrees to Eq. (\ref{vel}), because the momentum is defined in relativity
as $\mbox{\boldmath{$p$}} = E \mbox{\boldmath{$u$}}$ and the new relative
velocities are equal to each other in the two transformations.
The energy in the primed coordinate system is calculated from the energy vector
as
\begin{equation}
E' = \mbox{\boldmath{$E$}}' \cdot \hat{h}'= E \cosh \vartheta +
{\bf v}' \cdot \mbox{\boldmath{$p$}}= E \cosh \vartheta
+ \hat{n} \cdot \mbox{\boldmath{$p$}} \sinh \vartheta .
\end{equation}
The energy-momentum relation which is an invariant quantity in momentum space
is also calculated as
\begin{equation}
m^2=E'^2 + \mbox{\boldmath{$p$}}' \cdot \mbox{\boldmath{$p$}}'= E^2 +
\mbox{\boldmath{$p$}} \cdot \mbox{\boldmath{$p$}},\label{mass}
\end{equation}
which is equal to its mass squared.
Using the new relative velocity the transformation of the energy and the
momentum
with usual vectors are calculated as
\begin{eqnarray}
& &E' = E \cosh \vartheta + \hat{n} \cdot \mbox{\boldmath{$p$}}
\sinh \vartheta, \nonumber \\
& &\hat{n}' \cdot \mbox{\boldmath{$p$}}' = E \sinh \vartheta
+ \hat{n} \cdot \mbox{\boldmath{$p$}} \cosh \vartheta, \nonumber \\
& &\hat{n}' \times (\hat{n}' \times \mbox{\boldmath{$p$}}') = \hat{n} \times
(\hat{n} \times \mbox{\boldmath{$p$}}), \label{eptrans}
\end{eqnarray}
where the last two equations give again the transverse and normal components
of the
velocity addition rule which are Eq. (\ref{veladdition}), if they are divided
by the primed energy, that is, the first equation.

In the unprimed coordinate system, the energy and the momentum of a moving
particle are represented with a hyperbolic angle as
$E=m \cosh \alpha = m/\sqrt{1-u^2}$ and $\mbox{\boldmath{$p$}} = m \hat{m}
\sinh \alpha= m\mbox{\boldmath{$u$}}/\sqrt{1-u^2}$, respectively, where $u$
is the magnitude of the velocity of the particle.  From the above
transformations
for the energy and the momentum, it is also possible to do so in the primed
coordinate system by
using the hyperbolic trigonometric identities in Appendix B.  Using
Eq. (\ref{coshlaw}) in Appendix B, the energy of the particle in the primed
coordinate system is calculated as
\begin{eqnarray}
E' &=& E \cosh \vartheta + \hat{n} \cdot \mbox{\boldmath{$p$}}
\sinh \vartheta \nonumber \\
&=& m \cosh \alpha \cosh \vartheta + m \hat{n} \cdot \hat{m} \sinh \alpha
\sinh \vartheta \nonumber \\
&=& m \cos (\alpha \hat{m} \ominus \vartheta \hat{n}) = m \cosh \beta
= {m \over{\sqrt{1 -u'^2}}},\label{energy}
\end{eqnarray}
where $u'$ is the magnitude of the velocity of the particle which is observed
in the primed coordinate system.  From
Eqs. (\ref{mass}) and (\ref{energy}), the momentum of the particle in the
primed coordinate system is calculated as
\begin{equation}
\mbox{\boldmath{$p$}}' = \sqrt{m^2 - E'^2}
= {m \mbox{\boldmath{$u$}}' \over{\sqrt{1 -u'^2}}} = m \hat{l}' \sinh \beta,
\label{primedmom}
\end{equation}
where $\hat{l}' \cdot \hat{l}' = -1$.
This equation agrees to the following calculation by using Eqs. (\ref{eptrans})
and (\ref{sinhlaw}) in Appendix B
\begin{eqnarray}
\hat{n}' \cdot \mbox{\boldmath{$p$}}'
&=& \hat{n}' \cdot \hat{l}' p' = E \sinh \vartheta
+ \hat{n} \cdot \mbox{\boldmath{$p$}} \cosh \vartheta \nonumber \\
&=& m \cosh \alpha \sinh \vartheta + m \hat{n} \cdot \hat{m} \cosh \vartheta
\sinh \alpha \nonumber \\
&=& m \hat{n} \cdot [\hat{m} \sinh \alpha \cosh \vartheta - \hat{n} \sinh
\vartheta \cosh \alpha + {\gamma - 1 \over{\gamma}}\hat{n} \times (\hat{n}
\times \hat{m}) \sinh \alpha] \nonumber \\
&=& m \hat{n} \cdot \sin (\alpha \hat{m} \ominus \vartheta \hat{n})
= m \hat{n} \cdot \hat{l} \sinh \beta = \hat{n}' \cdot(m \hat{l}' \sinh \beta),
\label{pangle}
\end{eqnarray}
where the identities $\hat{n} \cdot \{ \hat{n} \times (\hat{n} \times \hat{m}) \}
= 0$, and
$\hat{n} \cdot \hat{n} = -1$ are used.  As previously mentioned, the direction
cosines of the unit vector of the relative velocity are invariant under the
Lorentz transformation, such as,
$\hat{n} \cdot \hat{i} = \hat{n}' \cdot \hat{i}' = - n_x$,
$\hat{n} \cdot \hat{j} = \hat{n}' \cdot \hat{j}' = - n_y$ and
$\hat{n} \cdot \hat{k} = \hat{n}' \cdot \hat{k}' = - n_z$.  This can be
generalized that the angle between two unit vectors is invariant under the
Lorentz transformation, that is,
$\hat{n}' \cdot \hat{l}' = \hat{n} \cdot \hat{l}$.
Hence the magnitude of the primed momentum is $p' = m \sinh \beta$.
Therefore the momentum vector of the particle in the primed coordinate system
agrees to Eq. (\ref{primedmom}).  The difference between the two unit vectors
$\hat{l}'$ and $\hat{l}$ is similar to that between the unit vectors
$\hat{n}'$ and $\hat{n}$ of the relative velocities.

Since the direction cosines
of an unit vector is invariant under the Lorentz transformation, the unit
vector is transformed as
\begin{eqnarray}
\hat{l}' &=& l_x \hat{i}' + l_y \hat{j}' + l_z \hat{k}' \nonumber \\
&=& \hat{l} + (\hat{n} \cdot \hat{l})(\hat{n} - \hat{n}'),
\end{eqnarray}
and thus, the angle between two unit vectors is calculated to be invariant
under the Lorentz
transformation as follows
\begin{eqnarray}
\hat{l}' \cdot \hat{n}' &=& [ \hat{l} - (\hat{n} \cdot \hat{l})(\hat{n}' -
\hat{n})] \cdot (\hat{n} \cosh \vartheta + \hat{h} \sinh \vartheta) \nonumber \\
&=& \hat{l} \cdot \hat{n}. \label{transdot}
\end{eqnarray}
More elegant form of the transformation of the unit vector can be written by
\begin{eqnarray}
& &\hat{l}' + (\hat{n}' \cdot \hat{l}')\hat{n}'
= \hat{l} + (\hat{n} \cdot \hat{l})\hat{n}, \nonumber \\
& &{\rm or}~~ \hat{n}' \times (\hat{n}' \times \hat{l}')
= \hat{n} \times (\hat{n} \times \hat{l}),\label{transcross}
\end{eqnarray}
which means that the normal component of the unit vector $\hat{l}$ with respect
to the unit
vector of the relative velocity is the same in both frames.
Therefore Eqs. (\ref{transdot}) and (\ref{transcross}) can be regarded as the
transformation rules for an unit vector.  The invariant quantity for an unit
vector is
\begin{equation}
\hat{l}' \cdot \hat{l}' = \hat{l} \cdot \hat{l} = -1,
\end{equation}
under the Lorentz transformation.  This is an inevitable condition for
relativity, because all the other inertial frames should have the same
measures, namely, units of time and length as the inertial frame where we live.

So using the transformation rules for an unit vector, the transverse and the
normal components of the momentum in the primed frame are written in
terms of the variables in the unprimed frame by
\begin{eqnarray}
& &\hat{n}' \cdot \mbox{\boldmath{$p$}}' = \cosh \vartheta [\hat{n} \cdot
\{ \mbox{\boldmath{$p$}} - E \mbox{\boldmath{$v$}} + {\gamma - 1 \over{\gamma}}
\hat{n} \times (\hat{n} \times \mbox{\boldmath{$p$}}) \} ],
\nonumber \\
& &\hat{n}' \times (\hat{n}' \times \mbox{\boldmath{$p$}}')
=  \cosh \vartheta[\hat{n} \times \{ \hat{n} \times (\mbox{\boldmath{$p$}}
- E \mbox{\boldmath{$v$}} + {\gamma - 1 \over{\gamma}} \hat{n} \times
(\hat{n} \times \mbox{\boldmath{$p$}})) \} ],
\end{eqnarray}
where the following equation obtained from Eq. (\ref{pangle}):
\begin{eqnarray}
p' \hat{l} &=&  \cosh \vartheta \{\mbox{\boldmath{$p$}}
- E \mbox{\boldmath{$v$}} + {\gamma - 1 \over{\gamma}} \hat{n} \times
(\hat{n} \times \mbox{\boldmath{$p$}}) \} \nonumber \\
&=& \mbox{\boldmath{$p$}} - (\gamma - 1) (\hat{n} \cdot
\mbox{\boldmath{$p$}})\hat{n}  - \gamma E \mbox{\boldmath{$v$}},\label{pfirst}
\end{eqnarray}
is used,
which is nothing but the component equation and agrees to the form of
Eq. (\ref{first}).
If these equations are divided by the primed energy, another form of the
velocity
addition rule is obtained as follows
\begin{eqnarray}
& &\hat{n}' \cdot \mbox{\boldmath{$u$}}' = {\hat{n} \cdot
\{ \mbox{\boldmath{$u$}} - \mbox{\boldmath{$v$}} + {\gamma - 1 \over{\gamma}}
\hat{n} \times (\hat{n}
\times \mbox{\boldmath{$u$}}) \} \over{1+ \mbox{\boldmath{$v$}}\cdot
\mbox{\boldmath{$u$}}}}= {\hat{n} \cdot
( \mbox{\boldmath{$u$}} - \mbox{\boldmath{$v$}}) \over{1+ \mbox{\boldmath{$v$}}\cdot
\mbox{\boldmath{$u$}}}}, \nonumber \\
& &\hat{n}' \times (\hat{n}' \times \mbox{\boldmath{$u$}}')
=  {\hat{n} \times [ \hat{n} \times \{ \mbox{\boldmath{$u$}}
- \mbox{\boldmath{$v$}} + {\gamma - 1 \over{\gamma}}\hat{n} \times
(\hat{n} \times \mbox{\boldmath{$u$}}) \}
] \over{1+ \mbox{\boldmath{$v$}}\cdot \mbox{\boldmath{$u$}}}}
=  {\hat{n} \times (\hat{n} \times \mbox{\boldmath{$u$}})
\over{\gamma (1+ \mbox{\boldmath{$v$}}\cdot \mbox{\boldmath{$u$}}})},
\end{eqnarray}
which should be compared with Eq. (\ref{veladdition}).
These equations agree to Eq. (\ref{veladdition}), if the irrelevant terms are
dropped out.  Since the two components
of the velocity in the primed coordinate system are described as the same form
in terms of the unprimed physical quantities, the magnitude of the
velocity of the particle in the primed coordinate system can be regarded as
\begin{equation}
u' = {|\mbox{\boldmath{$u$}} - \mbox{\boldmath{$v$}} + {\gamma - 1 \over{\gamma}}
\hat{n} \times (\hat{n}
\times \mbox{\boldmath{$u$}})| \over{1+ \mbox{\boldmath{$v$}}\cdot
\mbox{\boldmath{$u$}}}}
= \sqrt{1-{(1 - v^2)(1 - u^2) \over{(1 + \mbox{\boldmath{$v$}} \cdot
\mbox{\boldmath{$u$}})^2}}} \le 1,
\end{equation}
where the velocity vector of the particle in the primed coordinate systems can
be expressed as this equation without the symbol of the absolute value.
Of course
this velocity vector is also not the real primed velocity of the particle, but
the vector which has unprimed unit vector like Eq. (\ref{pfirst}).

The scalar product of two different
physical quantities is invariant under the Lorentz transformation, though their
new relative velocities are different from each other.
The new relative velocity in the coordinate transformation is different
from that in the momentum transformation, it is not difficult to calculate the
invariant quantity like followings
\begin{equation}
p_{\mu}' s'^{\mu} = E' t' + \mbox{\boldmath{$p$}}' \cdot \mbox{\boldmath{$r$}}'
= E t + \mbox{\boldmath{$p$}} \cdot \mbox{\boldmath{$r$}} = p_{\mu} s^{\mu},
\end{equation}
where the following calculations are used:
\begin{eqnarray}
& & {\bf v}_r \cdot {\bf v}_p tE = (\hat{n} \cdot \mbox{\boldmath{$r$}}) (\hat{n}
\cdot \mbox{\boldmath{$p$}}) \sinh^2 \vartheta - tE \sinh^2 \vartheta, \nonumber
\\
& & {\bf v}_r \cdot \mbox{\boldmath{$p$}} t = t (\hat{n}
\cdot \mbox{\boldmath{$p$}}) \cosh \vartheta \sinh \vartheta + (\hat{n} \cdot
\mbox{\boldmath{$r$}}) (\hat{n} \cdot \mbox{\boldmath{$p$}})
\sinh^2 \vartheta, \nonumber \\
& & {\bf v}_p \cdot \mbox{\boldmath{$r$}} E = E (\hat{n}
\cdot \mbox{\boldmath{$r$}}) \cosh \vartheta \sinh \vartheta + (\hat{n} \cdot
\mbox{\boldmath{$r$}}) (\hat{n} \cdot \mbox{\boldmath{$p$}})
\sinh^2 \vartheta,
\end{eqnarray}
where subscripts mean the kind of new relative velocities.

Like infinitesimal displacement vectors, infinitesimal displacement of the
momentum vectors are transformed as
\begin{eqnarray}
d\mbox{\boldmath{$p$}}' &=& d\mbox{\boldmath{$p$}} - {\bf v} dE, \nonumber \\
d\mbox{\boldmath{$E$}}' &=& d\mbox{\boldmath{$E$}} + {\bf v} dE,
\end{eqnarray}
where the new relative velocity is calculated as
\begin{eqnarray}
{\bf v} dE &=& {\bf v}' dE' = \hat{n} \cdot d\mbox{\boldmath{$p$}}
\sinh \vartheta \hat{h}' + dE \sinh \vartheta \hat{n}' \nonumber \\
&=& \hat{n}' \cdot d\mbox{\boldmath{$p$}}'
\sinh \vartheta \hat{h} + dE' \sinh \vartheta \hat{n}.
\end{eqnarray}
The scalar products of $\hat{h}$ and $\hat{n}$ to the new relative velocity
generate the following two Lorentz transformation equations and the double
vector products of the primed unit vector of the relative velocity to the
infinitesimal momentum displacement vector can be
calculated as
\begin{eqnarray}
& & dE' = dE \cosh \vartheta + \hat{n} \cdot d\mbox{\boldmath{$p$}}
\sinh \vartheta, \nonumber \\
& & \hat{n}' \cdot d\mbox{\boldmath{$p$}}' = dE \sinh \vartheta
+ \hat{n} \cdot d\mbox{\boldmath{$p$}} \cosh \vartheta, \nonumber \\
& & \hat{n}' \times(\hat{n}' \times d\mbox{\boldmath{$p$}}')
= \hat{n} \times (\hat{n} \times d\mbox{\boldmath{$p$}}).
\end{eqnarray}
Since a force is defined as the time derivative of a momentum, the
transformation of the force in vector form is calculated as
\begin{eqnarray}
\mbox{\boldmath{$F$}}' &=& {d\mbox{\boldmath{$p$}}' \over{dt'}}
={d\mbox{\boldmath{$p$}} - {\bf v}dE
\over{\gamma dt -  {\bf v}' \cdot d \mbox{\boldmath{$r$}}}}\nonumber \\
&=& {\mbox{\boldmath{$F$}} - {\bf v}{dE \over{dt}} \over{\gamma (1 +
{\mbox{\boldmath{$v$}}} \cdot \mbox{\boldmath{$u$}})}}
= {\mbox{\boldmath{$F$}} - {\bf v}{dE \over{dt}} \over{\gamma (1 +
\mbox{\boldmath{$v$}} \cdot \mbox{\boldmath{$p$}}/E)}}.
\end{eqnarray}
The power is transformed as
\begin{equation}
P' = {dE' \over{dt'}}={{dE \over{dt}} + {\mbox{\boldmath{$v$}}} \cdot
\mbox{\boldmath{$F$}} \over{1 + \mbox{\boldmath{$v$}}
\cdot \mbox{\boldmath{$u$}}}}={P + {\mbox{\boldmath{$v$}}} \cdot
\mbox{\boldmath{$F$}} \over{1 + \mbox{\boldmath{$v$}}
\cdot \mbox{\boldmath{$u$}}}}.
\end{equation}
The parallel and perpendicular components of the force to the relative velocity
are transformed as
\begin{eqnarray}
& & \hat{n}' \cdot \mbox{\boldmath{$F$}}' = {\hat{n} \cdot \mbox{\boldmath{$F$}}
- v {dE \over{dt}} \over{1 + \mbox{\boldmath{$v$}} \cdot \mbox{\boldmath{$u$}}
}}, \nonumber \\
& & \hat{n}' \times(\hat{n}' \times \mbox{\boldmath{$F$}}')
= {\hat{n} \times (\hat{n} \times \mbox{\boldmath{$F$}})
\over{\gamma(1 + \mbox{\boldmath{$v$}} \cdot \mbox{\boldmath{$u$}})}}.
\end{eqnarray}

Since derivatives of coordinates are transformed like a four vector, the
separate vector
forms of them can be written by
\begin{eqnarray}
\Box &=& {\partial \over{\partial t}} \check{h}
+ {\partial \over{\partial x}} \check{i}
+ {\partial \over{\partial y}} \check{j}
+ {\partial \over{\partial z}} \check{k}
= {\partial \over{\partial t}} \check{h} + \check{\nabla} \nonumber \\
&=&{\partial \over{\partial t}} \hat{h}
- {\partial \over{\partial x}} \hat{i}
- {\partial \over{\partial y}} \hat{j}
- {\partial \over{\partial z}} \hat{k}
= {\partial \over{\partial t}} \hat{h} - \nabla,
\end{eqnarray}
where the checked unit vectors mean contravariant vectors.  Since we are more
familiar with a gradient operator with covariant unit vectors than
contravariant ones, the minus sign should appear before the gradient
operator.
Using chain rule the vector forms of the derivatives transformation are
obtained as
\begin{eqnarray}
{\bf \nabla}' &=& {\bf \nabla} + {\bf v}
{\partial \over{\partial t}}, \nonumber \\
{\partial \over{\partial t'}} \hat{h}' &=& {\partial \over{\partial t}}
\hat{h} + {\bf v} {\partial \over{\partial t}},
\end{eqnarray}
where all the unit vectors are used with covariant vectors, because it is easy
to calculate scalar and vector products with other physical quantities which
almost have covariant unit vectors.
The new relative velocity in the transformation of derivatives is calculated
as
\begin{eqnarray}
{\bf v} {\partial \over{\partial t}} &=& {\bf v}' {\partial \over{\partial t'}}
= - \hat{h} \sinh \vartheta (\hat{n}' \cdot \nabla')
+ \hat{n} \sinh \vartheta {\partial \over{\partial t'}} \nonumber \\
&=& - \hat{h}' \sinh \vartheta (\hat{n} \cdot \nabla)
+ \hat{n}' \sinh \vartheta {\partial \over{\partial t}},
\end{eqnarray}
where the relation $\check{n} \cdot \check{\nabla} = \hat{n} \cdot \nabla$ is
needless to say.
The scalar products of $\hat{h}$ and $\hat{n}$ to the new relative velocity
and the double vector products of $\hat{n}'$ to the primed gradient
give
\begin{eqnarray}
& & {\partial \over{\partial t'}} = \cosh \vartheta{\partial \over{\partial t}}
- \sinh \vartheta \hat{n} \cdot \nabla, \nonumber \\
& & \hat{n}' \cdot \nabla' = \cosh \vartheta \hat{n} \cdot \nabla
- \sinh \vartheta {\partial \over{\partial t}}, \nonumber \\
& & \hat{n}' \times (\hat{n}' \times \nabla') = \hat{n} \times
(\hat{n} \times \nabla).
\end{eqnarray}
The invariant quantity for the derivatives is the D'Alembertian:
\begin{equation}
\Box'^2={\partial^2 \over{\partial t'^2}}-\nabla'^2
= {\partial^2 \over{\partial t^2}}-\nabla^2 = \Box^2,
\end{equation}
where the D'Alembertian is used with the squared quantity, because it is
necessary
to distinguish it from the four dimensional gradient operator.

If a charge and a current densities together transform like a four vector under
the Lorentz transformation, they can be written in vector form by
\begin{eqnarray}
& & \mbox{\boldmath{$j$}}' = \mbox{\boldmath{$j$}} - \rho {\bf v}, \nonumber \\
& & \mbox{\boldmath{$\rho$}}' = \mbox{\boldmath{$\rho$}} + \rho {\bf v},
\end{eqnarray}
where the new relative velocity for the charge and current densities can be
calculated as
\begin{eqnarray}
\rho {\bf v} &=& \rho' {\bf v}' = \hat{n} \cdot \mbox{\boldmath{$j$}}
\sinh \vartheta \hat{h}' + \rho \sinh \vartheta \hat{n}' \nonumber \\
&=& \hat{n}' \cdot \mbox{\boldmath{$j$}}'
\sinh \vartheta \hat{h} + \rho' \sinh \vartheta \hat{n}.
\end{eqnarray}
The scalar products of the unit vectors $\hat{h}$ and $\hat{n}$ to
the new relative velocity and the double vector products of $\hat{n}'$ to the
primed current vector give
\begin{eqnarray}
& & \rho' = \rho \cosh \vartheta + \hat{n} \cdot \mbox{\boldmath{$j$}}
\sinh \vartheta, \nonumber \\
& &\hat{n}' \cdot \mbox{\boldmath{$j$}}' = \rho \sinh \vartheta
+ \hat{n} \cdot \mbox{\boldmath{$j$}} \cosh \vartheta, \nonumber \\
& &\hat{n}' \times (\hat{n}' \times \mbox{\boldmath{$j$}}') = \hat{n} \times
(\hat{n} \times \mbox{\boldmath{$j$}}).
\end{eqnarray}
The invariant quantity for the charge and current densities can be calculated as
\begin{equation}
J'^{\mu}J'_{\mu} = \rho'^2 + \mbox{\boldmath{$j$}}' \cdot
\mbox{\boldmath{$j$}}' = \rho'^2 (1 - u'^2) = \rho^2 (1 - u^2) = \rho^2
+ \mbox{\boldmath{$j$}} \cdot \mbox{\boldmath{$j$}} = J^{\mu}J_{\mu},
\end{equation}
where the velocities in the parentheses
mean $\mbox{\boldmath{$u$}}' = \mbox{\boldmath{$j$}}'/\rho'$
and $\mbox{\boldmath{$u$}} = \mbox{\boldmath{$j$}}/\rho$.
Using the two kinds of the transformation equations for the derivatives and the
charge-current density the divergence of a current can be calculated to be an
invariant quantity
as
\begin{equation}
\partial'_{\mu}J'^{\mu} = \partial'_0 \rho' - \nabla' \cdot
\mbox{\boldmath{$j$}}' = \partial_0 \rho
- \nabla \cdot \mbox{\boldmath{$j$}} = \partial_{\mu}J^{\mu},
\end{equation}
where this continuity equation also can be obtained from the total derivative
of a charge density with respect to time as
\begin{equation}
{d \over{dt}} \rho = {\partial \rho \over{\partial t}}
+ {\partial \rho \over{\partial x}}{dx \over{dt}}
+ {\partial \rho \over{\partial y}}{dy \over{dt}}
+ {\partial \rho \over{\partial z}}{dz \over{dt}}= \partial_0 \rho
- \nabla \cdot \mbox{\boldmath{$j$}}.
\end{equation}
Therefore if the charge density is time independent in an inertial frame,
we know from the above two equations that the charge is conserved in all the
other inertial frames.

As shown above, if a scalar and a vector potentials together transform like a
four vector under the Lorentz transformation, which can be thought to be come
from the following covariant propagator
\begin{equation}
A^{\mu}(r) = \int d^4r' D(r - r')J^{\mu}(r'),
\end{equation}
where the propagator $D(r - r')$ should satisfy the wave equation for the
electromagnetic field, then its transformation property is the same as the
charge-current density.  The vector forms of the scalar and the vector
potentials
are written by
\begin{eqnarray}
& & \mbox{\boldmath{$A$}}' = \mbox{\boldmath{$A$}} - \phi {\bf v}, \nonumber \\
& & \mbox{\boldmath{$\phi$}}' = \mbox{\boldmath{$\phi$}} + \phi {\bf v},
\end{eqnarray}
where the new relative velocity is
\begin{eqnarray}
\phi {\bf v} &=& \phi' {\bf v}' = \hat{n} \cdot \mbox{\boldmath{$A$}}
\sinh \vartheta \hat{h}' + \phi \sinh \vartheta \hat{n}' \nonumber \\
&=& \hat{n}' \cdot \mbox{\boldmath{$A$}}'
\sinh \vartheta \hat{h} + \phi' \sinh \vartheta \hat{n}.
\end{eqnarray}
The scalar products of the unit vectors $\hat{h}$ and $\hat{n}$ to
the new relative velocity and the double vector products of $\hat{n}'$ to the
primed vector potential give
\begin{eqnarray}
& & \phi' = \phi \cosh \vartheta + \hat{n} \cdot \mbox{\boldmath{$A$}}
\sinh \vartheta, \nonumber \\
& &\hat{n}' \cdot \mbox{\boldmath{$A$}}' = \phi \sinh \vartheta
+ \hat{n} \cdot \mbox{\boldmath{$A$}} \cosh \vartheta, \nonumber \\
& &\hat{n}' \times (\hat{n}' \times \mbox{\boldmath{$A$}}') = \hat{n} \times
(\hat{n} \times \mbox{\boldmath{$A$}}).
\end{eqnarray}
The divergence of the scalar and vector
potentials is an invariant quantity under the Lorentz transformation which
is nothing but the Lorentz gauge:
\begin{equation}
\partial'_{\mu}A'^{\mu} = \partial'_0 \phi' - \nabla' \cdot
\mbox{\boldmath{$A$}}' = \partial_0 \phi
- \nabla \cdot \mbox{\boldmath{$A$}} = \partial_{\mu}A^{\mu}.
\end{equation}

The square of the sum or subtraction of two different physical quantities is
invariant under
the Lorentz transformation, if the dimensions and the transformation properties
of the two physical quantities are equal to each other.  As shown above the
energy-momentum and the scalar-vector potential have the same transformation
properties.  So the multiplication of the electromagnetic coupling constant,
which is a scalar under the Lorentz transformation, to the scalar and vector
potentials have the same dimensions as the energy-momentum relation.  The sums
of the two physical quantities transform as follows
\begin{eqnarray}
& &E'- e \phi' = (E - e \phi) \cosh \vartheta + (\hat{n} \cdot
\mbox{\boldmath{$p$}} - e \hat{n} \cdot \mbox{\boldmath{$A$}}) \sinh \vartheta,
\nonumber \\
& &\mbox{\boldmath{$p$}}' - e \mbox{\boldmath{$A$}}' = \mbox{\boldmath{$p$}}
- e \mbox{\boldmath{$A$}} - {\bf v}_p E + e {\bf v}_A \phi,
\end{eqnarray}
where the sum of the two new relative velocities is
\begin{eqnarray}
E {\bf v}_p - e \phi {\bf v}_A &=& \{(\hat{n} \cdot \mbox{\boldmath{$p$}} -
e \hat{n} \cdot \mbox{\boldmath{$A$}}) \hat{h}' + (E - e \phi) \hat{n}' \}
\sinh \vartheta  \nonumber \\
&=&\{ (\hat{n}' \cdot \mbox{\boldmath{$p$}}' - e \hat{n}' \cdot
\mbox{\boldmath{$A$}}')\hat{h} + (E' - e \phi')\hat{n} \} \sinh \vartheta.
\end{eqnarray}
Using the new relative velocity the following quantity is calculated to be
invariant under the Lorentz transformation as
\begin{eqnarray}
& &(E'- e \phi')^2 + (\mbox{\boldmath{$p$}}' - e \mbox{\boldmath{$A$}}') \cdot
(\mbox{\boldmath{$p$}}' - e \mbox{\boldmath{$A$}}') \nonumber \\
& & =(E- e \phi)^2 + (\mbox{\boldmath{$p$}} - e \mbox{\boldmath{$A$}}) \cdot
(\mbox{\boldmath{$p$}} - e \mbox{\boldmath{$A$}}) = M^2,
\end{eqnarray}
which corresponds that an equation of motion for a particle in the
electromagnetic fields is invariant under the Lorentz transformation, if the
invariant quantity is set to be the mass squared of the particle as shown in
gauge theories.

The $\gamma$ matrices used in the Dirac equation are transformed like the basis
vectors under the Lorentz transformation, because $(\gamma^0)^2 = 1$ and
$(\gamma^i)^2 = - 1$.  Using the direction cosines of the relative velocity the
$\gamma$ matrix for the direction of the relative velocity can be defined as
\begin{equation}
\gamma_n = n_x \gamma_1 + n_y \gamma_2 + n_z \gamma_3.
\end{equation}
The Lorentz transformation for the $\gamma$ matrices is similar to the basis
vectors as follows
\begin{eqnarray}
& &\gamma'_0 = \gamma_0 \cosh \vartheta + \gamma_n \sinh \vartheta, \nonumber \\
& &\gamma'_n = \gamma_n \cosh \vartheta + \gamma_0 \sinh \vartheta.
\end{eqnarray}
The $\gamma$ matrices for the space parts are transformed like an unit vector as
follows
\begin{eqnarray}
& &\gamma'_i \cdot \gamma'_n = \gamma_i \cdot \gamma_n, \nonumber \\
& &\gamma'_i = \gamma_i + (\gamma_i \cdot \gamma_n)(\gamma_n - \gamma'_n).
\end{eqnarray}
The scalar product of the $\gamma$ matrices to the above sum of the two
different physical quantities is invariant as follows
\begin{equation}
\gamma'^0(E'- e \phi')+ \mbox{\boldmath{$\gamma$}}' \cdot (\mbox{\boldmath{$p$}}'
- e \mbox{\boldmath{$A$}}') = \gamma^0(E- e \phi)+ \mbox{\boldmath{$\gamma$}}
\cdot (\mbox{\boldmath{$p$}} - e \mbox{\boldmath{$A$}}).
\end{equation}
This equation is not proved, but rather defined as all the other polar vectors
in this paper in a sense that the basis vectors are
chosen in the primed coordinate system in order to have the same direction
cosines of the primed
relative velocity as those of the unprimed relative velocity.  Therefore there
is no new relative velocity for the basis vectors.  Like the above calculations
the invariance of the square of sum of several energy-momentum relations is
useful in the energy momentum conservation for many physical reaction processes
\cite{yongkyu}.

Hitherto polar vectors are investigated under the Lorentz transformation, they
all have the similar transformation properties.  It is an interesting question
whether an axial vector has the same transformation property or not.
An angular momentum
vector is the very axial vector.  Since the definition of an angular momentum
vector consists of a position vector and a momentum vector for which we know
well
the Lorentz transformation properties, by using the
prescription for a vector product and the transformation matrices for the
coordinate and the momentum, an angular momentum in the primed frame is
calculated
as
\begin{eqnarray}
& &\mbox{\boldmath{$L$}}' = \mbox{\boldmath{$r$}}' \times \mbox{\boldmath{$p$}}'
\nonumber \\
& &= (y' p_z' - z' p_y') \hat{i}' + (z' p_x' - x' p_z') \hat{j}'
+ (x'p_y' - y'p_x') \hat{k}' \nonumber \\
& &= \mbox{\boldmath{$L$}} + (\hat{n} \cdot \mbox{\boldmath{$L$}})(\hat{n}
- \hat{n}') + (\hat{n} \times \mbox{\boldmath{$p$}})(\hat{n} \cdot
\mbox{\boldmath{$r$}}) - (\hat{n} \times \mbox{\boldmath{$r$}})(\hat{n} \cdot
\mbox{\boldmath{$p$}})- (\hat{n} \times \mbox{\boldmath{$p$}})(\hat{n}' \cdot
\mbox{\boldmath{$r$}}') + (\hat{n} \times \mbox{\boldmath{$r$}})(\hat{n}' \cdot
\mbox{\boldmath{$p$}}') \nonumber \\
& &= \mbox{\boldmath{$L$}} + (\hat{n} \cdot \mbox{\boldmath{$L$}})(\hat{n}
- \hat{n}') + \hat{n} \times (\hat{n} \times \mbox{\boldmath{$L$}})
- \hat{n}' \times (\hat{n}' \times \mbox{\boldmath{$L$}}'),
\end{eqnarray}
where it is easy to calculate the transformation for $\hat{n}' \cdot
\mbox{\boldmath{$L$}}' =\hat{n} \cdot \mbox{\boldmath{$L$}}$ by replacing the
basis vectors in the above equation with the direction cosines of $\hat{n}'$.
At first view the angular momentum seems to be transformed like an unit
vector under the
Lorentz transformation, but dose not so because it is difficult to confirm
that $\hat{n}' \times (\hat{n}' \times \mbox{\boldmath{$L$}}') = \hat{n} \times
(\hat{n} \times \mbox{\boldmath{$L$}})$.  The best way in this situation is to
calculate the transformation property of the magnitude of the angular momentum:
\begin{equation}
L'^2 = - \mbox{\boldmath{$L$}}' \cdot \mbox{\boldmath{$L$}}'
= - \mbox{\boldmath{$L$}} \cdot \mbox{\boldmath{$L$}}
+(E \mbox{\boldmath{$r$}} - t \mbox{\boldmath{$p$}})^2
- (E' \mbox{\boldmath{$r$}}' - t' \mbox{\boldmath{$p$}}')^2,\label{magang}
\end{equation}
where the following invariant relation:
\begin{equation}
\hat{n}' \cdot (E' \mbox{\boldmath{$r$}}' - t' \mbox{\boldmath{$p$}}')
= \hat{n} \cdot (E \mbox{\boldmath{$r$}} - t \mbox{\boldmath{$p$}}),
\end{equation}
is used, which is similar to the equation $\hat{n}' \cdot \mbox{\boldmath{$L$}}'
= \hat{n} \cdot \mbox{\boldmath{$L$}}$.  The magnitude of an angular
momentum is not an invariant quantity in relativity, but invariant with
$(E\mbox{\boldmath{$r$}} - t \mbox{\boldmath{$p$}})^2$.  So it is necessary to
calculate the transformation property of this physical quantity as follows
\begin{eqnarray}
& & E' \mbox{\boldmath{$r$}}' - t' \mbox{\boldmath{$p$}}' \nonumber \\
& & = (E \mbox{\boldmath{$r$}} - t \mbox{\boldmath{$p$}}) \cosh \vartheta
- \hat{n} \cdot (E \mbox{\boldmath{$r$}} - t \mbox{\boldmath{$p$}}) \sinh
\vartheta \hat{h} + \hat{n} \times \mbox{\boldmath{$L$}} \sinh \vartheta.
\label{et}
\end{eqnarray}
This calculation shows some hints on the transformation property for the
angular momentum, but not satisfactory.  Since a double cross product
physical quantity can be represented as scalar product physical quantities,
the following quantity for an angular momentum can be transformed properly
under the Lorentz transformation as
\begin{eqnarray}
& & \hat{n}' \times \mbox{\boldmath{$L$}}' = \hat{n}' \times
(\mbox{\boldmath{$r$}}'  \times \mbox{\boldmath{$p$}}')
= - \mbox{\boldmath{$r$}}'(\hat{n}' \cdot \mbox{\boldmath{$p$}}')
+ \mbox{\boldmath{$p$}}'(\hat{n}' \cdot \mbox{\boldmath{$r$}}')\nonumber \\
& & = \hat{n} \times \mbox{\boldmath{$L$}} \cosh \vartheta
+ \hat{n} \times [\hat{n} \times (E \mbox{\boldmath{$r$}}
- t \mbox{\boldmath{$p$}})] \sinh \vartheta,
\end{eqnarray}
where the last term has the following transformation property:
\begin{eqnarray}
& &\hat{n}' \times [\hat{n}' \times (E' \mbox{\boldmath{$r$}}'
- t' \mbox{\boldmath{$p$}}')] \nonumber \\
& & = \hat{n} \times [\hat{n} \times (E \mbox{\boldmath{$r$}}
- t \mbox{\boldmath{$p$}})] \cosh \vartheta
+ \hat{n} \times \mbox{\boldmath{$L$}} \sinh \vartheta.
\end{eqnarray}
It is interesting that this transformation should be compared with Eq.
(\ref{et}).  Thus the above two transformation equations and the invariance
of the transverse components of the angular momentum and
$E \mbox{\boldmath{$r$}}
- t \mbox{\boldmath{$p$}}$ are regarded as the
transformation rules for the angular momentum.
Therefore the invariant quantity for the angular momentum is
\begin{eqnarray}
& & (\hat{n}' \times \mbox{\boldmath{$L$}}')^2 - [\hat{n}' \times \{\hat{n}'
\times (E' \mbox{\boldmath{$r$}}' - t' \mbox{\boldmath{$p$}}') \} ]^2
\nonumber \\
& & = (\hat{n} \times \mbox{\boldmath{$L$}})^2 - [\hat{n} \times \{\hat{n}
\times (E \mbox{\boldmath{$r$}} - t \mbox{\boldmath{$p$}}) \} ]^2
\nonumber \\
& & = \mbox{\boldmath{$L$}} \cdot \mbox{\boldmath{$L$}}
+ (\hat{n} \cdot \mbox{\boldmath{$L$}})^2
- (E \mbox{\boldmath{$r$}} - t \mbox{\boldmath{$p$}}) \cdot
(E \mbox{\boldmath{$r$}} - t \mbox{\boldmath{$p$}})
- \{\hat{n} \cdot (E \mbox{\boldmath{$r$}} - t \mbox{\boldmath{$p$}}) \}^2
\end{eqnarray}
where the squared quantities mean the scalar product of the vectors in the
first two lines.  This agrees to Eq. (\ref{magang}), if the transverse
components are dropped out.

From the transformation property of the angular momentum, the transformation
for a torque can be calculated as
\begin{eqnarray}
& &\hat{n}' \cdot \mbox{\boldmath{$\tau$}}' = \hat{n}' \cdot
{d\mbox{\boldmath{$L$}}' \over{dt'}}
= {\hat{n} \cdot
\mbox{\boldmath{$\tau$}} \over{\gamma (1 + \mbox{\boldmath{$v$}}
\cdot \mbox{\boldmath{$u$}})}}, \nonumber \\
& & \hat{n}' \times \mbox{\boldmath{$\tau$}}'
= \hat{n}' \times {d\mbox{\boldmath{$L$}}' \over{dt'}}
= {\hat{n} \times \mbox{\boldmath{$\tau$}} + \hat{n} \times \{P
(\mbox{\boldmath{$v$}} \times \mbox{\boldmath{$r$}})
-t (\mbox{\boldmath{$v$}} \times \mbox{\boldmath{$F$}}) \}
\over{1 + \mbox{\boldmath{$v$}}
\cdot \mbox{\boldmath{$u$}}}}.
\end{eqnarray}
It is inevitable to include the counterpart of the angular momentum in the
transformation rules for the torque:
\begin{eqnarray}
& &\hat{n}' \cdot (P' \mbox{\boldmath{$r$}} - t' \mbox{\boldmath{$F$}}')
= {\hat{n} \cdot (P \mbox{\boldmath{$r$}} - t \mbox{\boldmath{$F$}})
\over{\gamma (1 + \mbox{\boldmath{$v$}} \cdot \mbox{\boldmath{$u$}})}},
\nonumber \\
& &\hat{n}' \times \{\hat{n}' \times (P' \mbox{\boldmath{$r$}}
- t' \mbox{\boldmath{$F$}}') \} = {\hat{n} \times \{\hat{n} \times
(P \mbox{\boldmath{$r$}} - t \mbox{\boldmath{$F$}}) +
\mbox{\boldmath{$v$}} \times \mbox{\boldmath{$L$}} \}
\over{1 + \mbox{\boldmath{$v$}} \cdot \mbox{\boldmath{$u$}}}}.
\end{eqnarray}

The electric and magnetic fields are the very axial vectors like an angular
momentum in electromagnetism, which are written by
\begin{equation}
\mbox{\boldmath{$B$}} = \nabla \times \mbox{\boldmath{$A$}},~~
\mbox{\boldmath{$E$}} = - \nabla \phi - {\partial \mbox{\boldmath{$A$}}
\over{\partial t}},
\end{equation}
where the difference between the electromagnetic fields and the angular
momentum is only the minus sign before the gradient operator due to its
covariant vector nature.
So they transform like an axial vector under the Lorentz transformation as
follows
\begin{eqnarray}
& & \hat{n}' \cdot \mbox{\boldmath{$B$}}'
= \hat{n} \cdot \mbox{\boldmath{$B$}}, \nonumber \\
& & \hat{n}' \cdot \mbox{\boldmath{$E$}}'
= \hat{n} \cdot \mbox{\boldmath{$E$}}, \nonumber \\
& & \hat{n}' \times \mbox{\boldmath{$B$}}'
= \hat{n} \times \mbox{\boldmath{$B$}} \cosh \vartheta
- \hat{n} \times (\hat{n} \times \mbox{\boldmath{$E$}}) \sinh \vartheta,
\nonumber \\
& & \hat{n}' \times (\hat{n}' \times \mbox{\boldmath{$E$}}')
= \hat{n} \times (\hat{n} \times \mbox{\boldmath{$E$}})\cosh \vartheta
- \hat{n} \times \mbox{\boldmath{$B$}} \sinh \vartheta.
\end{eqnarray}
This transformation shows that the origin of a magnetic field is due to the
Lorentz transformation \cite{Lorrain}, and is consistent with the Biot-Savart
law.
From this transformation property, the invariant quantity for the
electromagnetic field can be obtained as
\begin{eqnarray}
& &(\hat{n}' \times \mbox{\boldmath{$B$}}')^2 - \{\hat{n}' \times
(\hat{n}' \times \mbox{\boldmath{$E$}}') \}^2
= (\hat{n} \times \mbox{\boldmath{$B$}})^2 - \{\hat{n} \times
(\hat{n} \times \mbox{\boldmath{$E$}}) \}^2 \nonumber \\
& &= \mbox{\boldmath{$B$}} \cdot \mbox{\boldmath{$B$}}
+ (\hat{n} \cdot \mbox{\boldmath{$B$}})^2
- \mbox{\boldmath{$E$}} \cdot \mbox{\boldmath{$E$}}
- (\hat{n} \cdot \mbox{\boldmath{$E$}})^2.
\end{eqnarray}
Since the transverse components of the electric and magnetic fields are invariant
under the Lorentz transformation, if they are dropped out, the remaining parts
are also invariant as follows
\begin{equation}
B'^2 - E'^2 = B^2 - E^2 = {1 \over2}F_{\mu \nu}F^{\mu \nu},
\end{equation}
which is known as the norm of electromagnetic 2-form or Faraday \cite{Misner}.
This appears also in the Lagrangian for an electromagnetic interaction as
a photon field.

\section{Geometrical aspects of the Lorentz transformation}
\begin{figure}[h]
\centerline{\epsfig{file=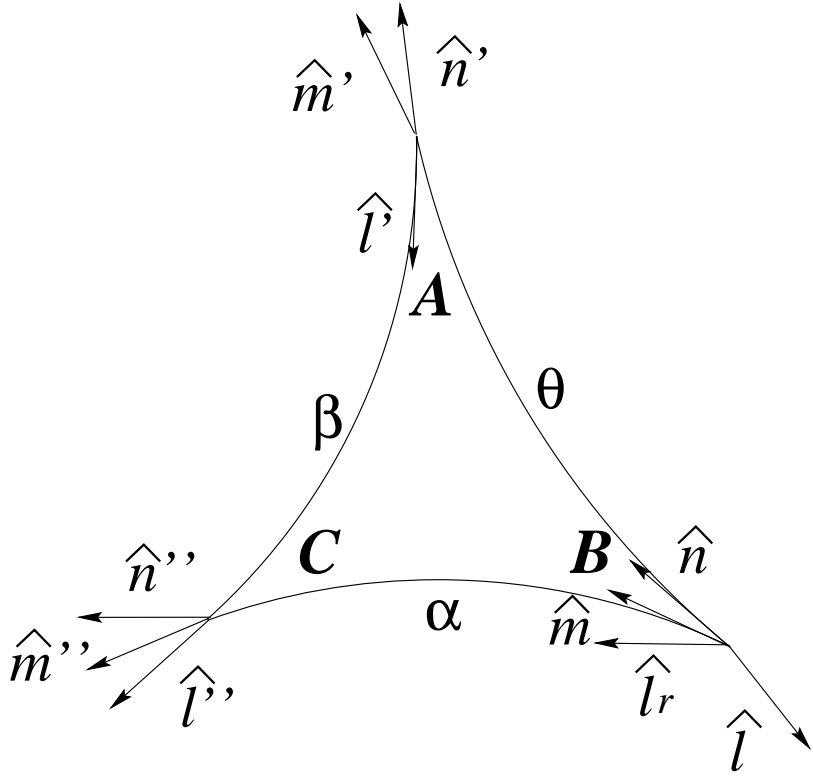, width=0.6\columnwidth}}
\caption{The three vertices of the hyperbolic triangle are the positions of the
origin of the primed frame, the origin of the unprimed frame and the particle
according to the angle $A$, $B$ and $C$.  The unit vector $\hat{l}_r$ can be
regarded as the complete rotated vector of $\hat{l}$ around the triangle. }
\label{fig2}
\end{figure}

The three angles $\alpha$, $\beta$, and $\vartheta$ mentioned for the
transformation in momentum space in the previous
section make a triangle in a hyperbolic space like Fig. \ref{fig2},
whose vertices are the
origins of the observing two frames $O$, and $O'$,
and the position of an observed
particle $P$.  While the three angles $\alpha$, $\beta$, and $\vartheta$ are
global parameters, the unit vectors $\hat{l}$, $\hat{m}$ and $\hat{n}$ are
defined locally at each vertex which are tangential directions on
a hyperbolic sphere.  So
these unit vectors are designated by unprimed vectors in the rest frame, primed
vectors in the moving frame and double primed vectors in the reference frame of
the particle.

An observer in the rest frame sees the origin of the moving
frame with the velocity $\mbox{\boldmath{$v$}}= v \hat{n}
= \hat{n} \tanh \vartheta $ and the moving particle with the momentum
$\mbox{\boldmath{$p$}} = p
\hat{m} = m \hat{m} \sinh \alpha$ whose velocity is $\mbox{\boldmath{$u$}} =
{{\mbox{\boldmath{$p$}}}/E}= \hat{m} \tanh \alpha$.  The angle between the
two velocities $\mbox{\boldmath{$v$}}$, and $\mbox{\boldmath{$u$}}$ can be
defined as $\cos B = - \hat{n} \cdot \hat{m}$.  An observer in the moving
frame sees the origin of the rest frame with the velocity
$\mbox{\boldmath{$v$}}' = - v \hat{n}' = - \hat{n}' \tanh \vartheta$ and the
moving particle with the momentum $\mbox{\boldmath{$p$}}' = p' \hat{l}'
= m \hat{l}' \sinh \beta$ whose velocity is $\mbox{\boldmath{$u$}}' =
{\mbox{\boldmath{$p$}}' /E'}= \hat{l}' \tanh \beta$.  The angle between
the two velocities $\mbox{\boldmath{$v$}}'$, and $\mbox{\boldmath{$u$}}'$ can
be defined as $\cos A = \hat{n}' \cdot \hat{l}'$.  Since the magnitude of the
relative velocity between two origins of reference frames is observed to be
the same in
both reference frames, an observer in the frame of the moving particle sees
the origin of the rest frame with the velocity $\mbox{\boldmath{$u$}}''_r
= - \hat{m}'' \tanh \alpha$ and the origin of the moving frame with the velocity
$\mbox{\boldmath{$u$}}''_m = - \hat{l}'' \tanh \beta$, respectively.  The angle
between the two velocities $\mbox{\boldmath{$u$}}''_r$ and
$\mbox{\boldmath{$u$}}''_m$ can be defined as $\cos C = - \hat{m}'' \cdot
\hat{l}''$.

The law of cosines in hyperbolic trigonometry shown in Appendix B can be
applied to this triangle as
\begin{eqnarray}
& & \cosh \alpha = \cosh \beta \cosh \vartheta - \hat{n}' \cdot \hat{l}'
\sinh \beta \sinh \vartheta,\nonumber \\
& & \cosh \beta = \cosh \alpha \cosh \vartheta + \hat{n} \cdot \hat{m}
\sinh \alpha \sinh \vartheta,\nonumber \\
& & \cosh \vartheta = \cosh \alpha \cosh \beta + \hat{m}'' \cdot \hat{l}''
\sinh \alpha \sinh \beta. \label{nlawofcos}
\end{eqnarray}
If the mass of the observed particle is multiplied to the above equations,
these equations are nothing but the Lorentz transformation as follows
\begin{eqnarray}
& & E = E' \cosh \vartheta - \hat{n}' \cdot \mbox{\boldmath{$p$}}' \sinh
\vartheta, \nonumber \\
& & E' = E \cosh \vartheta + \hat{n} \cdot \mbox{\boldmath{$p$}} \sinh
\vartheta, \nonumber \\
& & m^4 \cosh \vartheta = E E'(m^2 + \mbox{\boldmath{$p$}} \cdot
\mbox{\boldmath{$p$}}') + \cosh \vartheta(p^2 p'^2- p^2 E'
\mbox{\boldmath{$v$}}' \cdot \mbox{\boldmath{$p$}}' + p'^2 E
\mbox{\boldmath{$v$}} \cdot \mbox{\boldmath{$p$}} ),
\end{eqnarray}
where the last equation is obtained from the calculation of $\cos C =
- \hat{m}'' \cdot \hat{l}''$ by using the transformation equations:
\begin{eqnarray}
& & \hat{m}'' = \hat{m} \cosh \alpha + \hat{h} \sinh \alpha, \nonumber \\
& & \hat{l}'' = \hat{l}' \cosh \beta + \hat{h}' \sinh \beta.
\end{eqnarray}
These transformations are explained again later.
Since the unit vector $\hat{l}''$ is the transformed vector from the primed
coordinate system, it is worth to note that $\hat{m}'' \cdot \hat{l}''=
\hat{m} \cdot \hat{l}_r \not= \hat{m} \cdot \hat{l}$ compared to
$\hat{n}' \cdot \hat{l}' =
\hat{n} \cdot \hat{l}$ in the previous section.
The reason is shown in Fig. \ref{fig2}.
Therefore the angle between two unit vectors is invariant under the Lorentz
transformation
between only two inertial frames.
The last equation is further simplified as
\begin{eqnarray}
& & (E-E')^2+(\mbox{\boldmath{$p$}}-\mbox{\boldmath{$p$}}')^2+2EE'(1-\cosh
\vartheta)=0, \nonumber \\
& & {\rm or}~~\mbox{\boldmath{$p$}} \cdot \mbox{\boldmath{$p$}}'
= - p^2 - \sinh \vartheta E' (\hat{n} \cdot \mbox{\boldmath{$p$}})
= \mbox{\boldmath{$p$}} \cdot (\mbox{\boldmath{$p$}}- {\bf v}E),
\end{eqnarray}
where the last equation agrees to Eq. (\ref{mo}) and is of importance to check
whether transformed formulas are correct or not.
The first two equations in the above Lorentz transformation give the remaining
partners of the Lorentz transformation
by inserting each of them into the other equation as follows
\begin{eqnarray}
& & \hat{n} \cdot \mbox{\boldmath{$p$}} = \hat{n}' \cdot \mbox{\boldmath{$p$}}'
\cosh \vartheta - E' \sinh \vartheta, \nonumber \\
& & \hat{n}' \cdot \mbox{\boldmath{$p$}}' = \hat{n} \cdot \mbox{\boldmath{$p$}}
\cosh \vartheta + E \sinh \vartheta.
\end{eqnarray}
Therefore the Lorentz transformation for an energy-momentum can be interpreted
geometrically as the law of cosines for the hyperbolic triangle.

The law of cosines for the corresponding polar triangle is applied to the
hyperbolic triangle as
\begin{eqnarray}
& & \cos A = - \cos B \cos C + \sin B \sin C \cosh \alpha, \nonumber \\
& & \cos B = - \cos A \cos C + \sin A \sin C \cosh \beta, \nonumber \\
& & \cos C = - \cos A \cos B + \sin A \sin B \cosh \vartheta.
\end{eqnarray}
If the cosines of the angles are replaced by the unit vectors defined above,
then the following
relations are obtained
\begin{eqnarray}
& &  \cos A = \hat{n}' \cdot \hat{l}' = \hat{n}'' \cdot \hat{l}''
-  (\hat{m}'' \times \hat{n}'') \cdot (\hat{m}'' \times \hat{l}'')
+ \sin B \sin C \cosh \alpha, \nonumber
\\
& & \cos B = - \hat{n} \cdot \hat{m} = - \hat{n}'' \cdot \hat{m}''
+  (\hat{l}'' \times \hat{m}'') \cdot (\hat{l}'' \times \hat{n}'')
+ \sin A \sin C \cosh \beta, \nonumber
\\
& & \cos C = - \hat{m}'' \cdot \hat{l}'' = - \hat{m} \cdot \hat{l}
+  (\hat{n} \times \hat{l}) \cdot (\hat{n} \times \hat{m})
+ \sin A \sin B \cosh \vartheta \nonumber \\
& & \hspace{3cm}= - \hat{m}' \cdot \hat{l}'
+  (\hat{n}' \times \hat{l}') \cdot (\hat{n}' \times \hat{m}')
+ \sin A \sin B \cosh \vartheta.
\end{eqnarray}
These equations are of importance for the relations among the three angles
$A,B$
and $C$ and play a crucial role for the derivation of the law of sine just
later.

There are three pairs of the Lorentz transformation among the three inertial
frames.  In
order to obtain another form of the law of sine for the hyperbolic
triangle, it is necessary to remind the Lorentz transformation
for the
unit vectors among the three inertial frames.
The Lorentz transformation for the unit vectors of the moving frames is again
written as shown in previous section by
\begin{eqnarray}
& & \hat{n}' = \hat{n} \cosh \vartheta + \hat{h} \sinh \vartheta, \nonumber \\
& & \hat{h}' = \hat{h} \cosh \vartheta + \hat{n} \sinh \vartheta.
\end{eqnarray}
The Lorentz transformation for the unit vectors of the reference frame of the
particle can be written in terms of the unit vectors of the moving frame by
\begin{eqnarray}
& & \hat{l}'' = \hat{l}' \cosh \beta + \hat{h}' \sinh \beta, \nonumber \\
& & \hat{h}'' = \hat{h}' \cosh \beta + \hat{l}' \sinh \beta,
\end{eqnarray}
because the relative velocity between the two frames is the velocity of the
particle observed in the moving frame.
The Lorentz transformation for the unit vectors of the reference frame of the
particle can be written in terms of the unit vectors of the rest frame by
\begin{eqnarray}
& & \hat{m}'' = \hat{m} \cosh \alpha + \hat{h} \sinh \alpha, \nonumber \\
& & \hat{h}'' = \hat{h} \cosh \alpha + \hat{m} \sinh \alpha,
\end{eqnarray}
because of the same reason above.
The unit vector of the time component for the reference frame of the particle
should be the same in the above two transformations.
So the following relation is obtained
\begin{equation}
\hat{h} \cosh \alpha + \hat{m} \sinh \alpha
= \hat{h}' \cosh \beta + \hat{l}' \sinh \beta,
\end{equation}
which means that every time axis of all inertial frames has the same nature
as that of the inertial frame where we live, though it is rotated through the
Lorentz transformation, as if the heavens look the same everywhere on the earth
in regards to a spherical space.  The reason can be thought that, while the
time axis is only one, the space axes are three so that the Lorentz
transformation causes mutual rotations among the space axes like the unit
vectors
$l_r$ and $l$ as shown in Fig. \ref{fig2}.

The scalar product of $\hat{h}$ to the identity yields
\begin{eqnarray}
\cosh \alpha &=& \hat{h} \cdot \hat{h}' \cosh \beta + \hat{h} \cdot \hat{l}'
\sinh \beta \nonumber \\
&=& \cosh \beta \cosh \vartheta - \cos A \sinh \beta \sinh \vartheta,
\end{eqnarray}
which is the law of cosine for the angle $\alpha$.
The scalar product of $\hat{h}'$ to the identity yields
\begin{eqnarray}
\cosh \beta &=& \hat{h}' \cdot \hat{h} \cosh \alpha + \hat{h}' \cdot \hat{m}
\sinh \alpha \nonumber \\
&=& \cosh \alpha \cosh \vartheta - \cos B \sinh \alpha \sinh \vartheta,
\end{eqnarray}
which is the law of cosine for the angle $\beta$.  These two laws of cosine
agree to Eq. (\ref{nlawofcos}).
The scalar product of $\hat{l}'$ to the identity gives
\begin{eqnarray}
\hat{l}' \cdot \hat{l}' \sinh \beta &=&  \hat{l}' \cdot \hat{h} \cosh \alpha
+ \hat{l}' \cdot \hat{m} \sinh \alpha \nonumber \\
&=& - \hat{l}' \cdot \hat{n}' \cosh \alpha \sin \vartheta
+ \hat{l}' \cdot \hat{m} \sinh \alpha.
\end{eqnarray}
If a mass is multiplied to the equation, the resulting equation is written by
\begin{equation}
\hat{l}' \cdot \mbox{\boldmath{$p$}}' = \hat{l}' \cdot (\mbox{\boldmath{$p$}} -
\hat{n}' E \sinh \vartheta) = \hat{l}' \cdot (\mbox{\boldmath{$p$}} - {\bf
v}E),
\end{equation}
where $- \hat{h}'(\hat{n} \cdot \mbox{\boldmath{$p$}}) \sinh \vartheta$
is added
because the space unit vector is orthogonal to the unit vector of time.  The
law of sine for the hyperbolic triangle means the Lorentz transformation for
a momentum vector, because the magnitude of a momentum can be
expressed as a hyperbolic sine.
This equation is further calculated as
\begin{eqnarray}
& & \hat{l}' \cdot \hat{l}' \sinh \beta =  \hat{l} \cdot \hat{l} \sinh \beta
\nonumber \\
& &= \hat{l} \cdot \hat{m} \cosh \vartheta \sinh \alpha- \hat{l} \cdot \hat{n}
\cosh \alpha \sinh \vartheta
- (\hat{n} \times \hat{m}) \cdot (\hat{n} \times \hat{l})(\cosh \vartheta -1)
\sinh \alpha,
\end{eqnarray}
where the law of cosines for the polar triangle and the following calculation:
\begin{equation}
\hat{l}' \cdot \hat{m} = (\hat{n} \times \hat{m}) \cdot (\hat{n} \times
\hat{l}) - (\hat{n} \cdot \hat{m})(\hat{n} \cdot \hat{m}) \cosh \vartheta,
\end{equation}
are used.  Then the hyperbolic sine is written again in vector form by
\begin{equation}
\hat{l} \sinh \beta
=\hat{m} \cosh \vartheta \sinh \alpha - \hat{n} \cosh \alpha \sinh \vartheta
+ \hat{n} \times (\hat{n} \times \hat{m}) (\cosh \vartheta -1) \sinh \alpha.
\end{equation}
This can be regarded as another form of the law of sine in hyperbolic
trigonometry compared to Eq. (\ref{sinlaw}) in spherical trigonometry.  So the
sum of the squares of the hyperbolic cosine and sine is equal to unity as follows
\begin{equation}
\cosh^2 \beta + \hat{l} \cdot \hat{l} \sinh^2 \beta =1.
\end{equation}
This also means the energy momentum relation, if a square of mass is multiplied
to the equation.
If a mass is multiplied to the law of sine, the result is the primed momentum
vector described in unprimed coordinate system
\begin{eqnarray}
p' \hat{l}&=& \cosh \vartheta (\mbox{\boldmath{$p$}} - E \mbox{\boldmath{$v$}})
+ \hat{n} \times (\hat{n} \times \mbox{\boldmath{$p$}}) (\cosh \vartheta -1),
\nonumber \\
&=& \mbox{\boldmath{$p$}} - (\gamma - 1) (\hat{n} \cdot
\mbox{\boldmath{$p$}})\hat{n}  - \gamma E \mbox{\boldmath{$v$}},\label{ppfirst}
\end{eqnarray}
as shown in the previous section.  Instead of the unprimed unit vectors
if all the unit vectors in the equation are replaced by primed unit vectors,
then this equation
means that the primed momentum is expressed in terms of the unprimed momentum
described in the primed coordinate system which is not a real unprimed momentum.
The real primed momentum is expressed in terms of the real unprimed momentum as
\begin{equation}
\mbox{\boldmath{$p$}}' = p' \hat{l}' = p' \{ \hat{l} + (\hat{n} \cdot \hat{l})
(\hat{n} - \hat{n}') \} = \mbox{\boldmath{$p$}} - {\bf v} E,
\end{equation}
where $p' \hat{l}$ is substituted by Eq. (\ref{ppfirst}).

The scalar product of $\hat{m}$ to the identity gives
\begin{eqnarray}
& & \sinh \alpha = - \hat{m} \cdot \hat{n} \sinh \vartheta \cosh \beta
- \hat{m} \cdot \hat{l}' \sinh \beta \nonumber \\
& &= - \hat{m}' \cdot \hat{n}' \cosh \beta \sinh \vartheta - \hat{m}'
\cdot \hat{l}' \cosh \vartheta \sinh \beta
+ (\hat{n}' \times \hat{m}') \cdot (\hat{n}' \times \hat{l}')(\cosh \vartheta -1)
\sinh \beta,
\end{eqnarray}
which gives the following relations in the same way above
\begin{eqnarray}
& &\hat{m} \cdot \mbox{\boldmath{$p$}} = \hat{m} \cdot (\mbox{\boldmath{$p$}}' +
{\bf v} E) =\hat{m} \cdot (\mbox{\boldmath{$p$}}' +
{\bf v}' E') , \nonumber \\
& &p \hat{m}' = \cosh \vartheta (\mbox{\boldmath{$p$}}'
+ E' \mbox{\boldmath{$v$}}')
+ \hat{n}' \times (\hat{n}' \times \mbox{\boldmath{$p$}}') (\cosh \vartheta -1).
\end{eqnarray}
These equations are the inverse transformation of the momentum.

As the same way the equalities of the time component unit vector in the primed
and unprimed coordinate systems:
\begin{eqnarray}
& &\hat{h}' = \hat{h} \cosh \vartheta + \hat{n} \sinh \vartheta = \hat{h}''
\cosh \beta - \hat{l}'' \sinh \beta, \nonumber \\
& &\hat{h} = \hat{h}' \cosh \vartheta - \hat{n}' \sinh \vartheta = \hat{h}''
\cosh \alpha - \hat{l}'' \sinh \alpha,
\end{eqnarray}
give similar law of cosines and sines by scalar products of suitable unit
vectors.

The following calculation gives interesting results
\begin{eqnarray}
\hat{m}'' \cdot \hat{l}'' &=& (\hat{m} \cosh \alpha + \hat{h} \sinh \alpha)
\cdot (\hat{l}' \cosh \beta + \hat{h}' \sinh \beta) \nonumber \\
&\not = & \hat{m} \cdot \hat{l}
\end{eqnarray}
which is not a contradiction to the invariance of angle under the Lorentz
transformation as mentioned previously, if we consider that the space time
is not flat.  This is
related to the Thomas precession, because successive Lorentz transformations
reduce to a pure Lorentz boost and a rotational transformation
\cite{Goldstein,Jackson}.
This rotational transformation can be represented as the unit vectors $l$
and $l_r$ in Fig. \ref{fig2}.
If the observed particle is accelerated, then the
angle between $l$ and $l_r$ which is the angle defect of the hyperbolic triangle
in the hyperbolic space
are also varying.  Therefore the angler velocity for the angle defect can be
obtained.  This may be the Thomas precession in this geometrical
interpretation.

A simple example is very illustrative for the hyperbolic triangle.  Let us
calculate all the angles involved in the triangle.
A moving particle with a momentum
$\mbox{\boldmath{$p$}} = p \hat{j}$ whose mass is detected as 1 MeV is observed
in the rest frame.  Another inertial frame is moving with a relative velocity
$\mbox{\boldmath{$v$}} = v \hat{i}$.  The energy and the momentum of the
particle are calculated to be observed in the moving frame as
$E' = E \cosh \vartheta$ and
$\mbox{\boldmath{$p$}}' = p' \hat{l}' = - E \sinh \vartheta \hat{i}'
+ p \hat{j}'$ by using the Lorentz transformation.
The three hyperbolic angles of the triangle are calculated as
\begin{eqnarray}
& &\tanh \alpha = u = {p \over{E}} = {p \over{\sqrt{1 + p^2}}}, \nonumber \\
& &\tanh \beta = u' = {p' \over{E'}} ={\sqrt{p^2 + E^2 \sinh^2 \vartheta}
\over{E \cosh \vartheta}} = \sqrt{u^2 + v^2 - u^2 v^2}, \nonumber \\
& &\tanh \vartheta = v.
\end{eqnarray}
The three real angles of the hyperbolic triangle are
\begin{eqnarray}
& &\cos A = - \hat{i} \cdot \hat{j} = 0, \nonumber \\
& &\cos B = \hat{l}' \cdot \hat{i}' = {E \sinh \vartheta \over{\sqrt{p^2
+ E^2 \sinh^2 \vartheta}}} = {v \over{\sqrt{u^2 + v^2 - u^2 v^2}}},\nonumber \\
& &\cos C = - \hat{l}'' \cdot \hat{j}'' = - \cos A \cos B + \sin A \sin B \cosh
\vartheta \nonumber \\
& &= {p \cosh \vartheta \over{\sqrt{p^2 + E^2 \sinh^2 \vartheta}}}
= {u \over{\sqrt{u^2 + v^2 - u^2 v^2}}}.
\end{eqnarray}
If $u = v$, the hyperbolic triangle is an equilateral right triangle so that
$\cos B = \cos C = 1/\sqrt{2- u^2}$.  Table \ref{table}
shows some numerical calculations
for this triangle from non-relativistic to relativistic regions.
\begin{table}[h]
\begin{ruledtabular}
\begin{tabular}{|c|c|c|c|c|c|c|}
$\tanh \vartheta$ & $\tanh \alpha$ & $\tanh \beta$ & $A$ & $B$ & $C$ &
$180-A-B-C$\\ \hline\hline
0.1 & 0.1 & 0.141067 & 90 & 44.856 & 44.856 & 0.28792\\
0.3 & 0.3 & 0.414608 & 90 & 43.6496 & 43.6496 & 2.7008\\
0.5 & 0.5 & 0.661438 & 90 & 40.8934 & 40.8934 & 8.21321\\
0.7 & 0.7 & 0.860174 & 90 & 35.5323 & 35.5323 & 18.9355\\
0.9 & 0.9 & 0.981784 & 90 & 23.5519 & 23.5519 & 42.8962\\
0.99 & 0.99 & 0.999802 & 90 & 8.02958 & 8.02958 & 73.9408\\
0.999 & 0.999 & 0.999998 & 90 & 2.56 & 2.56 & 84.88\\
\end{tabular}
\end{ruledtabular}
\caption{\label{table} The three sides and the three angles of the hyperbolic
triangle are calculated from non-relativistic to relativistic regions.  The
angle defect runs from zero to $90~^o$ according to the relativistic motion
of the observed particle and the relative velocity between two inertial frames.}
\end{table}

The triangle of Fig. \ref{fig2} is quite different from a triangle in a
plane whose sum
of the three angles is $180~^o$.  The sum of the three angles for a hyperbolic
triangle is less than  $180~^o$, while the sum of the three angles for a
spherical triangle is more than  $180~^o$.  The general equation for the sum of
the three angles for a triangle is
\begin{equation}
A+B+C = \pi \pm {{\rm area} \over{R^2}},
\end{equation}
where the area means the area of the triangle and $R$ means the radius of a
sphere (+) or a hyperbolic sphere ($-$).
The minus sign is due to the imaginary radius of the hyperbolic sphere
\cite{Laptev,Kells,Coxeter}.  The curvature $R$ is infinite in plane
trigonometry.  The more
relativistic are the motion of an observed particle and the relative velocity,
the larger is the area of the hyperbolic
triangle as shown in table \ref{table}.
The maximum value of the angle defect is
approaching to $90~^o$ in this example.  For an equilateral triangle the
maximum value of the angle defect is approaching to $180~^o$, which can be
easily shown with the law of cosine because all the three angles and the
hyperbolic angles, respectively, are equal.  The angle between the relative
velocities is expressed with the velocity as $\cos A = 1/(1 + \sqrt{1-v^2})$.

\section{Conclusions}
In this paper we have studied the separate notation of the vector forms of the
Lorentz transformation and its geometrical aspects.  The four vector
notation have indeed powerful merits not only in the special theory of
relativity but also in the general theory of relativity for the covariance of
physical laws \cite{Misner}.  However the detailed structure of space time is
not seen in the four vector notation.  The separate notation of four vectors
can show some clues for the geometry of our universe, besides it can express
the vector form of the Lorentz transformation as covariant forms.
Three inertial frames
whose relative velocities among them are all non-zero make a hyperbolic
triangle.  The sum of its three angles is less than $180~^o$.  While the space
of an inertial frame is flat which is a tangential space on a hyperbolic sphere,
the space of all inertial frames is a hyperbolic space, namely, a Lobatchewsky
space which are on the hyperbolic sphere.  It is more easy to imagine this
picture on an unit sphere rather than a hyperbolic sphere, if some results are
taken into account carefully, such as, the imaginary radius of the hyperbolic
sphere, metric tensor, and so on.

Naturally two inertial frames are rotated mutually with a hyperbolic angle with
respect to each other.  Therefore the vector equations of the Lorentz
transformation between the inertial frames in the separate notation of
four vectors
can not be represented without mixing time and space components.  Such mixed
quantities are expressed in the new relative velocity.  The $\gamma$ factor
appeared in the component equations of the Lorentz transformation between
two frames does not
appears in the vector equations, because their own unit vectors are used.
This factor comes from the scalar product between the unit vectors of the two
frames in the calculation of components from the vector transformation
equation.
It is shown that we can
do with these vector equations what we can do with the components of the
Lorentz transformation, such as, calculations of invariant quantities.  It is
possible to transform an axial vector in the four dimension under the Lorentz
transformation which are represented as a tensor in the component form of the
Lorentz transformation.

No matter how this paper is worthless, at least very adventurous astronauts who
want to journey around the universe should know this work, if they want to
return home where their grandsons or grand-grandsons will welcome them due to
the time dilation of their space ship.  Navigators on the sea should know
that the earth is round so that their direction for the destination should be
determined by spherical trigonometry according to their sailed distance.
However navigators in the universe should know that the space is not flat so
that their direction for the destination should be determined by hyperbolic
trigonometry according to the velocity of their space ship.

\appendix
\section{angle vector}
An angle has its magnitude and unique direction to be able to define. However
it is not treated as a proper vector in the standard
texts \cite{Goldstein,Marion},
on the contrary infinitesimal angles, angular velocities and angular
accelerations are regarded as
proper vectors.  This treatment for an angle encounters a contradiction, when an
angular velocity summed with two vectors or decomposed into several vectors is
integrated with respect to time.  This integration gives a finite angle vector.
Therefore a finite angle should be defined
also as a proper vector as the angular velocity and acceleration are done.  The
problems to make it difficult to define an angle as a vector may be the
definition of the sum of two angle vectors or confusion with a rotational
operation.  Using spherical trigonometry, such problems are resolved.

\begin{figure}
\centerline{\epsfig{file=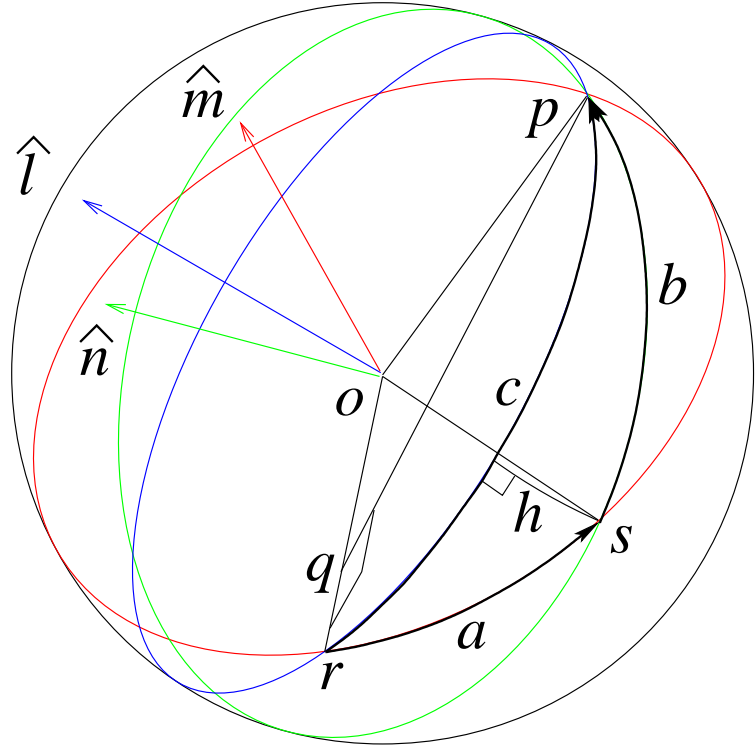, width=0.6\columnwidth}}
\caption{An angle vector can be defined as an arc on an unit sphere like
this figure.
The  angle vectors $a \hat{m}, b \hat{n}$ and $c \hat{l}$ have the arcs $a, b$
and $c$ as their magnitudes and the unit vectors $\hat{m}, \hat{n}$ and
$\hat{l}$ as their directions, respectively.
The figure shows how to add angle vectors, that is, $c\hat{l} = a \hat{m}
\oplus b \hat{n}$.  The length $\overline{oq}$ is
equal to $\cos c$, and the length $\overline{pq}$ is $\sin c$.
}
\label{fig3}
\end{figure}

To begin with, let us define an angle vector as a least arc arrow
on an unit sphere which is naturally lying on the great circle of the sphere
as shown in Fig. \ref{fig3}.
The magnitude of the angle vector is the length of the arc and its
direction is represented as an arrow on the sphere, or equivalently an unit
vector at the
center of the sphere perpendicular to the plane which is made of the center of
the sphere and the arc arrow according to the right handed rule.

If two angle vectors are equivalent, then the
magnitudes of the two angle vectors are equal to each other and they lie on the
same great circle on the sphere with the same direction of an arrow.  The
negative angle vector is represented as the arc arrow with opposite direction
of the
angle vector.

The sine and cosine of an angle vector have the following relations:
\begin{equation}
\cos \alpha \hat{n} = \cos \alpha,~~~\sin \alpha \hat{n} = \hat{n} \sin \alpha,
\end{equation}
which can be checked by expanding them.  If an angle vector is the sum of two
angle vectors as shown in Fig. \ref{fig3}:
\begin{equation}
c \hat{l} = a \hat{m} \oplus b \hat{n},
\end{equation}
where it is necessary to distinguish the addition of angle vectors from that
of usual
vectors so that the different type of an addition symbol should be employed,
then the sum of the angle vectors satisfies the following relations:
\begin{eqnarray}
& &\cos c = \cos (a \hat{m} \oplus b \hat{n}) = \cos a \cos b - \hat{m} \cdot
\hat{n} \sin a \sin b,\label{coslaw} \\
& &\hat{l} \sin c = \sin (a \hat{m} \oplus b \hat{n}) = \hat{m} \sin a \cos b +
\hat{n}\cos a  \sin b - \hat{m} \times \hat{n} \sin a \sin b.\label{sinlaw}
\end{eqnarray}
Therefore the magnitude and the direction of the angle vector $c \hat{l}$ are
calculated as
\begin{eqnarray}
& & c= \cos^{-1} [ \cos a \sin b + \hat{m} \cdot
\hat{n} \sin a \sin b], \nonumber \\
& &\hat{l} ={1\over{\sin c}}[ \hat{m} \sin a \cos b +
\hat{n}\cos a  \sin b - \hat{m} \times \hat{n} \sin a \sin b],
\end{eqnarray}
and thus the angle vector addition rule is neither commutative nor anti-commutative.
It is very important to keep the order of operations for angle vectors.  The
calculation for other angle vector, as an example,
should be $b \hat{n} = - a \hat{m}
\oplus c \hat{l}$.

The proof of the above relations is just referring to spherical trigonometry
\cite{Kells}.  A
spherical triangle on an unit sphere has three sides, $a,b$ and $c$, and the
corresponding opposite angles $A,B$ and $C$, respectively, as shown in
Fig. \ref{fig3}.
All the three sides
lie on their great circles.  Each angle included by two sides is the angle
between the sectional planes of the great circles on which the sides lie.
The laws of sines and cosines for the spherical triangle are
\begin{eqnarray}
& & {\sin a \over{\sin A}} = {\sin b \over{\sin B}} = {\sin c \over{\sin C}},
\\
& & \cos a = \cos b \cos c + \sin b \sin c \cos A,\label{cosa} \\
& & \cos b = \cos a \cos c + \sin a \sin c \cos B,\label{cosb} \\
& & \cos c = \cos a \cos b + \sin a \sin b \cos C.\label{cosc}
\end{eqnarray}
If we take $\hat{m} \cdot \hat{n}= - \cos C$ in Eq. (\ref{cosc}),
then Eq. (\ref{coslaw}) is proved.

If we calculate the laws of cosine as $[{\rm Eq. (\ref{cosc})} \times
\cos a + {\rm Eq. (\ref{cosb})}]/\sin a$ and $[{\rm Eq. (\ref{cosc})} \times
\cos b + {\rm Eq. (\ref{cosa})}]/\sin b$, then we get the following relations:
\begin{eqnarray}
& & \cos A \sin c = \cos a \sin b - \cos C \sin a \cos b, \\
& & \cos B \sin c = \sin a \cos b - \cos C \cos a \sin b.
\end{eqnarray}
These equations agree to the scalar products of unit vectors $\hat{m}$
and $\hat{n}$ to Eq. (\ref{sinlaw}), respectively, if we take
$\hat{l} \cdot \hat{n}=  \cos A$ and $\hat{l} \cdot \hat{m}=  \cos B$.  The
last term of Eq. (\ref{sinlaw}) can be put in by hand considering the unity
of sine squared and cosine squared in the right and left handed sides,
but it can be also proved.
The sum of the above equations is calculated as
\begin{eqnarray}
\sin c &=& {1- \cos C \over{\cos A + \cos B}} \cos a \sin b
+ {1- \cos C \over{\cos A + \cos B}}\sin a  \cos b \nonumber \\
&=& \cos A \cos a \sin b + \cos B \sin a \cos b + \sin B \sin C \sin^2 a \sin
b, \label{comsinc}
\end{eqnarray}
where the law of cosines for the polar triangle is used, which are the same
as that for the spherical triangle except the sign.  The polar triangle
corresponding to the above spherical triangle has the three angles $A',B'$
and $C'$
and the corresponding sides $a',b'$ and $c'$ which are related to that of
the spherical
triangle as follows
\begin{eqnarray}
A' &=& \pi - a,~~~B' = \pi - b,~~~C' = \pi - c, \nonumber \\
a' &=& \pi - A,~~~b' = \pi - B,~~~c' = \pi - C,
\end{eqnarray}
then the following relations are satisfied
\begin{eqnarray}
& & \cos A = - \cos B \cos C + \sin B \sin C \cos a,\label{pcosa} \\
& & \cos B = - \cos A \cos C + \sin A \sin C \cos b,\label{pcosb} \\
& & \cos C = - \cos A \cos B + \sin A \sin B \cos c.\label{pcosc}
\end{eqnarray}
This polar triangle can be imagined in Fig. \ref{fig3}
as the triangle made by the three
end points of the unit vectors $- \hat{l}, \hat{m}$ and $\hat{n}$, where the
minus sign before $\hat{l}$ is due to the angle c which is not cyclic here.
The above equation (\ref{comsinc}) is nothing but the scalar product of
$\hat{l}$ to Eq. (\ref{sinlaw}), so it is understood that $\hat{l} \cdot
\hat{m} \times \hat{n} = - \sin a \sin B \sin C$, where $- \sin C$ comes
from the
cross product of the unit vectors $\hat{m}$ and $\hat{n}$, and $\sin a \sin B$
which is equal to $\sin b \sin A$ comes from the dot product of the two
vectors $\hat{l}$ and $\hat{m} \times \hat{n}$ by using the Napier's rule for
the right spherical triangle, which is made by drawing the arc $h$ perpendicular
to the side $c$ from the vertex at which the angle $C$ is.  The arc $h$ is
the complementary angle to the angle between the two vectors $\hat{l}$ and
$\hat{m} \times \hat{n}$, so the Napier's rule for the arc $h$ is written by
\begin{equation}
\cos ({\pi \over{2}} - h) = \sin h = \sin a \sin B = \sin b \sin A.
\end{equation}

Since the angular velocities of the three angles are defined as
\begin{equation}
\mbox{\boldmath{$\omega$}}_a = {d(a\hat{m}) \over{dt}},~~~
\mbox{\boldmath{$\omega$}}_b = {d(b\hat{n}) \over{dt}},~~~
\mbox{\boldmath{$\omega$}}_c = {d(c\hat{l}) \over{dt}},
\end{equation}
the time derivative of $\cos c$ is calculated as
\begin{eqnarray}
- \sin c {dc \over{dt}} &=& - (\sin a \cos b + \hat{n} \cdot \hat{m}
\cos a \sin b) {da \over{dt}} \nonumber \\
&-& (\hat{n} \cdot \hat{m} \sin a \cos b + \cos a \sin b) {db \over{dt}}.
\end{eqnarray}
This equation can be rewritten by
\begin{eqnarray}
\hat{l} \sin c {d(c \hat{l}) \over{dt}} &=& (\hat{m} \sin a \cos b + \hat{n}
\cos a \sin b - \hat{m} \times \hat{n} \sin a \sin b) {d(a \hat{m}) \over{dt}}
\nonumber \\
&+& (\hat{m} \sin a \cos b + \hat{n} \cos a \sin b - \hat{m} \times \hat{n}
\sin a \sin b) {d(b \hat{n}) \over{dt}},
\end{eqnarray}
which is just the sum of angular velocities
\begin{equation}
\mbox{\boldmath{$\omega$}}_c = \mbox{\boldmath{$\omega$}}_a
\oplus \mbox{\boldmath{$\omega$}}_b.
\end{equation}
This agrees to the direct time derivative of
$c \hat{l} = a \hat{m} \oplus b \hat{n}$.
This equation shows that the angular velocity regarded as a proper vector
in the standard texts has the same properties as the angle vector defined
above.
The angular acceleration is also calculated in the same way as the angular
velocity is done.  Considering that the angular velocities of the rotation and
the precession for a symmetrical top can not be interchanged, non-commutative
properties of the rotations as mentioned in the standard texts
\cite{Goldstein,Marion} may not be a proper reason for the insufficient
definition of an angle vector.
As the same way, we can not rotate a symmetrical top
after precessing by using an angular acceleration, that is, a non-rotating
symmetrical top does not precess.  This angle vector addition rule also
appears partly in the matrix for the Eulerian angles in the rigid body motion.

\section{Hyperbolic trigonometry}
It is difficult to imagine a hyperbolic space directly which is known as
a Lobatchewsky space \cite{Coxeter}.  So the hyperbolic space is supposed
to be the space on an unit sphere as shown in the previous section with an
imaginary radius which can be called a hyperbolic sphere here.
If the metric for the space part is taken as $\hat{n} \cdot \hat{n}=-1$,
then the sine and cosine
of the angle vector are written by hyperbolic sine and cosine as follows
\begin{equation}
\cos \alpha \hat{n} = \cosh \alpha,~~~\sin \alpha \hat{n} = \hat{n} \sinh
\alpha.
\end{equation}
Due to the metric the sum of two hyperbolic angle vectors is expressed with
hyperbolic sines and cosines as
\begin{eqnarray}
& &\cosh c = \cos (a \hat{m} \oplus b \hat{n}) = \cosh a \cosh b + \hat{m}
\cdot \hat{n} \sinh a \sinh b,\label{coshlaw} \\
& &\hat{l} \sinh c = \sin (a \hat{m} \oplus b \hat{n}) = \hat{m} \sinh a
\cosh b + \hat{n} \cosh a \sinh b - i\hat{m} \times \hat{n} \sinh a
\sinh b,
\end{eqnarray}
where the last equation should have an imaginary term in order that the
cosine squared and sine squared should be unity.
Since the last term is not imaginary in the scalar equation as
\begin{equation}
\sinh c
= \cos A \cosh a \sinh b + \cos B \sinh a \cosh b
- \sin B \sin C \sinh^2 a \sinh
b, \label{comsinhc}
\end{equation}
which is also derived from the laws of sines and cosines for a hyperbolic
triangle and its polar triangle as shown below.  Therefore, in principle, it may
be possible to derive another form of law of sine in vector form as follows
\begin{eqnarray}
\hat{l} \sinh c &=& \sin (a \hat{m} \oplus b \hat{n}) \nonumber \\
&=&\hat{m} \sinh a \cosh b + \hat{n} \sinh b \cosh a
+ \hat{n} \times (\hat{n} \times \hat{m}) (\cosh b -1) \sinh a,\label{sinhlaw}
\end{eqnarray}
which is obtained in section IV in a different way as shown in Appendix A.

The unit
vectors for the directions of the hyperbolic angle vectors are defined
globally, as shown in the spherical space,
in the hyperbolic space where the triangle is.
It is difficult to calculate the operations between such unit vectors because
we don't know how to draw the unit vectors in the hyperbolic space.  Instead of
treating the unit vectors in such a way, it is much more proper and easy to
treat the unit vectors as local variables than global ones at the starting and
end points of the hyperbolic angle vectors.  Since a velocity in the Lorentz
transformation can be represented as a hyperbolic angle, the directions of
the relative velocities in two inertial frames are
defined as the unit vectors at the starting point and the end point of the
hyperbolic angle
on the hyperbolic sphere.
Therefore the unit vectors defined locally in the hyperbolic space are well
interpreted as
physical quantities and treated easily.

The law of sines and cosines for a hyperbolic triangle are
\begin{eqnarray}
& & {\sinh a \over{\sin A}} = {\sinh b \over{\sin B}} = {\sinh c \over{\sin C}},
\\
& & \cosh a = \cosh b \cosh c - \sinh b \sinh c \cos A,\label{cosha} \\
& & \cosh b = \cosh a \cosh c - \sinh a \sinh c \cos B,\label{coshb} \\
& & \cosh c = \cosh a \cosh b - \sinh a \sinh b \cos C.\label{coshc}
\end{eqnarray}
The law of cosines for the polar triangle corresponding to the hyperbolic
triangle are
\begin{eqnarray}
& & \cos A = - \cos B \cos C + \sin B \sin C \cosh a,\label{pcosha} \\
& & \cos B = - \cos A \cos C + \sin A \sin C \cosh b,\label{pcoshb} \\
& & \cos C = - \cos A \cos B + \sin A \sin B \cosh c.\label{pcoshc}
\end{eqnarray}

\end{document}